%% file: distributed_SVD_journal.tex
\documentclass[journal,onecolumn]{IEEEtran}
\usepackage{amsmath,amssymb,amsfonts,bm}
\usepackage{setspace}
\usepackage{graphicx}
\usepackage{textcomp}
\usepackage{highlight,xcolor}
\usepackage{colortbl}
\usepackage{bbding}
\usepackage{enumitem}
\usepackage{tblr-extras}
\UseTblrLibrary{caption}
\usepackage{cases}
\usepackage{subcaption}
\usepackage{makecell}
\usepackage{multirow}

\input{util_def}

\newcommand\copyrighttext{%
	\footnotesize \textcopyright This work has been submitted to the IEEE for possible publication. Copyright may be transferred without notice, after which this version may no longer be accessible.
}
\makeatletter
\def\ps@IEEEtitlepagestyle{
	\def\@oddfoot{\mycopyrightnotice}
	\def\@evenfoot{}
}
\def\mycopyrightnotice{
	{\footnotesize
		\begin{minipage}{\textwidth-2\fboxsep}%
			\centering%
			\noindent\fbox{\parbox{\linewidth}{\copyrighttext}}
		\end{minipage}
	}
}

\usepackage{nccmath}

\newtheorem{remark}{Remark}
  \usepackage{pgfplots}
  \usepackage{pgfplotstable}
  \pgfplotsset{compat=newest}
  \usetikzlibrary{plotmarks}
  \usetikzlibrary{arrows.meta}
  \usepgfplotslibrary{patchplots}
  \usepackage{grffile}
  \usetikzlibrary{spy,backgrounds,fit}
  \usetikzlibrary{calc}
  
\usepackage{tikz}
\usepackage[allcolors=blue,colorlinks=true]{hyperref}
\usepackage{orcidlink}

\newcommand{\myvspace}{\vspace{0pt}}

\newcommand{\tr}[1]{\operatorname{tr}\left ( #1 \right )}

\newcommand{\diag}[1]{\operatorname{diag}\left( #1 \right)}
\newcommand{\ps}[1]{\operatorname{nc}\left( #1 \right)}
\newcommand{\nc}[2]{\operatorname{NC}_{#1}\left( #2 \right)}

\newtheorem{theorem}{Theorem}

\begin{document}
\pagestyle{empty}

\title{Decentralized Singular Value Decomposition for Large-scale Distributed Sensor Networks}


\author{\IEEEauthorblockN{{Yufan Fan\orcidlink{0000-0003-3611-0191},~\IEEEmembership{Student Member,~IEEE}, and Marius Pesavento\orcidlink{0000-0003-3395-2588},~\IEEEmembership{Senior Member,~IEEE}}}
	\thanks{This work was financially supported by the Federal Ministry of Education and Research of Germany in the project Open6GHub under Grant 16KISK014. An earlier version of this work was presented in part at the 2024 IEEE International Conference on Acoustics, Speech and Signal Processing (ICASSP), Seoul, Korea, Republic of, April 14-19, 2024 [DOI: 10.1109/ICASSP48485.2024.10447748]. (\textit{Corresponding author: Yufan Fan.})}
	\thanks{Y. Fan and M. Pesavento are with the Communication Systems Group, Technische Universit\"at Darmstadt, 64283 Darmstadt, Germany (e-mail: yufan.fan@nt.tu-darmstadt.de; pesavento@nt.tu-darmstadt.de).}
	\thanks{\copyrighttext}
}

\maketitle

\thispagestyle{empty}


\begin{abstract}
This article studies the problem of decentralized Singular Value Decomposition (d-SVD), which is fundamental in various signal processing applications. Two scenarios are considered depending on the availability of the data matrix under consideration. In the first scenario, the matrix of interest is row-wisely available in each local node in the network. In the second scenario, the matrix of interest implicitly forms an outer product from two different series of measurements. By combining the lightweight local rational function approximation approach with parallel averaging consensus algorithms, two d-SVD algorithms are proposed to cope with the two aforementioned scenarios. We evaluate the proposed algorithms using two application examples: decentralized sensor localization via low-rank matrix completion and decentralized passive radar detection. Moreover, a novel and non-trivial truncation technique, which employs a representative vector that is orthonormal to the principal signal subspace, is proposed to further reduce the communication cost associated with the d-SVD algorithms. Simulation results show that the proposed d-SVD algorithms converge to the centralized solution with reduced communication cost compared to those facilitated with the state-of-the-art decentralized power method.
\end{abstract}


\input{intro}

\input{pre}

\input{dSVDknown}

\input{dSVDunknown}

\input{simulation}
\section{Conclusion}
	\label{sec: conclusion}
	In this article, we propose decentralized implementations of the SVD based on efficient local rational function approximation approach and parallel averaging consensus algorithms in two scenarios depending on the availability of the matrix of interest. First, the matrix of interest is row-wisely accessible to the nodes in the network. Second, the matrix of interest implicitly forms an outer product from two distinct series of measurements that are distributively available in the network. The proposed d-SVD algorithms are successfully applied to decentralized sensor localization and decentralized passive radar detection that are associate with the two aforementioned scenarios, respectively. All simulation results show that the solution of our d-SVD algorithms converge to the centralized solution at a reduced total communication cost compared to algorithms utilizing the d-PM. Furthermore, the proposed non-trivial truncation technique significantly reduces the communication cost of the d-SVD algorithms by tracking only the principal signal subspace, which is useful, e.g., in PCA and CCA applications where dominant singular values are present.
\appendices
\input{appendixRA}
\input{appendix1}

\bibliographystyle{IEEEtran}
\bibliography{references.bib}

\end{document}

%% file: util_def.tex
\definecolor{light-gray}{gray}{0.65}

\newcommand\ba{\ensuremath{\boldsymbol{a}}}

\newcommand\be{\ensuremath{\boldsymbol{e}}}

\newcommand\bn{\ensuremath{\boldsymbol{n}}}
\newcommand\bp{\ensuremath{\boldsymbol{p}}}
\newcommand\bq{\ensuremath{\boldsymbol{q}}}
\newcommand\br{\ensuremath{\boldsymbol{r}}}
\newcommand\bs{\ensuremath{\boldsymbol{s}}}
\newcommand\bu{\ensuremath{\boldsymbol{u}}}
\newcommand\bv{\ensuremath{\boldsymbol{v}}}
\newcommand\bw{\ensuremath{\boldsymbol{w}}}
\newcommand\bx{\ensuremath{\boldsymbol{x}}}
\newcommand\by{\ensuremath{\boldsymbol{y}}}
\newcommand\bz{\ensuremath{\boldsymbol{z}}}

\newcommand\bA{\ensuremath{\boldsymbol{A}}}
\newcommand\bB{\ensuremath{\boldsymbol{B}}}
\newcommand\bC{\ensuremath{\boldsymbol{C}}}

\newcommand\bG{\ensuremath{\boldsymbol{G}}}
\newcommand\bH{\ensuremath{\boldsymbol{H}}}
\newcommand\bI{\ensuremath{\boldsymbol{I}}}

\newcommand\bM{\ensuremath{\boldsymbol{M}}}

\newcommand\bK{\ensuremath{\boldsymbol{K}}}

\newcommand\bQ{\ensuremath{\boldsymbol{Q}}}
\newcommand\bR{\ensuremath{\boldsymbol{R}}}

\newcommand\bU{\ensuremath{\boldsymbol{U}}}
\newcommand\bV{\ensuremath{\boldsymbol{V}}}
\newcommand\bW{\ensuremath{\boldsymbol{W}}}
\newcommand\bX{\ensuremath{\boldsymbol{X}}}
\newcommand\bY{\ensuremath{\boldsymbol{Y}}}

\newcommand\bzero{\ensuremath{\boldsymbol{0}}}
\newcommand\bone{\ensuremath{\boldsymbol{1}}}

\newcommand\bphi{\ensuremath{\boldsymbol{\phi}}}

\newcommand\bkappa{\ensuremath{\boldsymbol{\kappa}}}

\newcommand\bTheta{\ensuremath{\boldsymbol{\Theta}}}
\newcommand\bLambda{\ensuremath{\boldsymbol{\Lambda}}}
\newcommand\bSigma{\ensuremath{\boldsymbol{\Sigma}}}
\newcommand\bPhi{\ensuremath{\boldsymbol{\Phi}}}

\newcommand{\norm}[1]{\lVert {#1} \rVert}

\newcommand{\abs}[1]{\lvert {#1} \rvert}

\newcommand{\floor}[1]{\lfloor{#1}\rfloor}

\newcommand{\tT}{\ensuremath{{\mathsf{T}}}}
\newcommand{\tH}{\ensuremath{{\mathsf{H}}}}
\newcommand{\tF}{\mathsf{F}}

%% file: intro.tex
\section{Motivation and Introduction}\label{sec:sysMod}

\IEEEPARstart{S}{ingular} Value Decomposition (SVD) serves as a fundamental matrix factorization technique that is valuable across various domains \cite{schmidtSurveysingular2020}, including environmental monitoring \cite{linApplicationSVD2019}, recommendation systems \cite{chenSurveycollaborative2018}, and Canonical Correlation Analysis (CCA) \cite{yangSurveycanonical2021}. In conventional scenarios, the SVD is typically performed centrally on data aggregated at a fusion center. However, this centralized approach faces significant challenges due to computational limitations and communication bottlenecks and becomes increasingly impractical in scaling to large-scale systems, such as modern Wireless Sensor Networks (WSNs), where measurements from hundreds or thousands of sensors need to be processed. These scalability issues necessitate alternative approaches for handling massive amounts of distributed sensor data. To this end, decentralized algorithms have emerged as a preferred solution. In the decentralized setup, the computational load is distributed across individual nodes - sensors with simple processing, storage, and communication capabilities. Nodes may consist of individual sensors or collections of sensors, such as sensor subarrays or clusters of sensor subarrays in WSNs. By enabling local computations and limiting information exchange to neighboring nodes, decentralized approaches effectively mitigate the scalability issues inherent in centralized processing.

The decentralized approaches can be categorized into two classes based on the availability of the measurements over the network, i.e., the sample-wise partitioning and the feature-wise partitioning \cite{wuReviewdistributed2018,gangDistributedprincipal2021}. In the sample-wise partitioning, each node has access to a different subset of samples of the dataset that contains the entire set of features. In contrast, in the feature-wise partitioning, each node has access to all observations of a single feature (or an exclusive subset of features). This arises naturally, for example, in distributed sensor deployments, where each individual sensor captures distinct features or aspects of a common object. In this article, the feature-wise partitioning is the focus, and is also referred to as the row-wise partitioning from the matrix point of view. Various decentralized SVD (d-SVD) algorithms for the feature-wise partitioning have been proposed in the literature \cite{liCommunicationefficientdistributed2020,liuPrivacypreservingfederated2023,guoFedPowerPrivacypreserving2024}. These algorithms typically operate in two stages: first, local computation on individual datasets, such as the renowned Power Method (PM), and second, solution aggregation. The aggregation phase may involve techniques such as orthogonal Procrustes transformation, either coordinated by a central unit or decentralized across the network. Apart from the PM, a d-SVD algorithm is proposed in \cite{hegedusFullydistributed2014} that combines the Stochastic Gradient Descent (SGD) approach and the gossiping learning framework \cite{ormandiGossiplearning2013} to compute the SVD iteratively. However, due to the nature of the SGD method, an appropriate step size must be chosen to achieve a balance between the convergence speed and the result accuracy.

The aforementioned d-SVD algorithms consider only the case where the data under consideration is directly partitioned in different local nodes. Nevertheless, in many applications, such as those relying on the CCA, the principal components of empirical cross-correlation matrices generated from two series of measurements need to be computed distributedly. Due to the extremely large data dimension, restrained transmission rate, and limited local processing resources, such matrices can not be extracted and directly partitioned. To this end, the Power SVD algorithm is proposed in \cite{wuDecentralizedarray2019,liParalleldistributed2021}, which combines the consensus algorithms and the decentralized PM (d-PM) \cite{scaglioneDecentralizedestimation2008}, and updates the left and right singular vectors interchangeably. Recently, a Distributed Adaptive Signal Fusion (DASF) framework has been proposed in \cite{musluogluUnifiedalgorithmic2023,musluogluUnifiedalgorithmic2023a}, where in each iteration a selected node receives compressed data from other nodes, solves a lower-dimensional version of the original network-wide problem, and shares its own compressed local solution with other nodes according to a specific network protocol.

Motivated by our previous work on the decentralized Eigenvalue Decomposition (d-EVD) \cite{fanDecentralizedeigendecomposition2021,fanDecentralizedeigendecomposition2023}, we propose in this article two types of d-SVD algorithms to cope with both aforementioned scenarios, i.e., the data matrix of interest is directly partitioned across nodes, and is represented as an outer product that is generated from two series of measurements. In our proposed d-SVD algorithms the data available at each node is diffused through the network using parallel consensus protocols with local interactions between nodes. Instead of the popular PM, our proposed d-SVD algorithms are based on the local lightweight Rational function Approximation (RA) approach that updates all singular values parallelly in an online manner. Different from the PM, the RA approach does not suffer from slow convergence if the singular values are closely separated or even have multiplicities. More generally, it does not suffer from divergence as encountered by other non power iteration based algorithms \cite{fanDecentralizedeigendecomposition2023}. At termination, each node has access to entire singular values and one row (or multiple rows) of the singular vector matrices corresponding to the node index. Since all eigenvalues are simultaneously computed in the RA approach, it is associated with a significant undesired overhead in large-scale sensor networks, e.g., for Principal Component Analysis (PCA) applications. To this end, we propose a novel and non-trivial truncation technique that only tracks the principal subspace, and the associated communication cost and the local computational overhead are significantly reduced.

We evaluate the performance of our d-SVD algorithms in two prominent application examples corresponding to the two aforementioned cases: (A) the decentralized sensor localization and (B) the decentralized passive radar detection. To summarize, our contributions are as follows:

\begin{itemize}[leftmargin=19pt]
	\item We address the d-SVD problem in two scenarios, where the data matrix of interest i) is partitioned in different nodes and ii) needs to be implicitly computed as an outer product from two series of measurements whose samples are distributively available in each node. Combining the lightweight local rational function approximation approach and parallel averaging consensus algorithms, two d-SVD algorithms are proposed to deal with the two scenarios respectively.
	\item To further reduce the overall communication cost and the local computational overhead, we propose a novel and non-trivial truncation technique, such that the proposed d-SVD algorithms track only the principal signal subspace, which is useful in PCA applications.
	\item We evaluate the two proposed d-SVD algorithms in important application examples, i.e., the decentralized sensor localization and the decentralized passive radar detection.
\end{itemize}

The article is organized as follows. In Section \ref{sec:pre} we introduce the problem formulation, and briefly explain and revisit the d-EVD algorithm that is based on the RA approach. Two d-SVD algorithms for the cases where the data matrix of interest is row-wisely partitioned and is implicitly formed as an outer product of two different series of measurements are proposed in Section \ref{sec:dSVDknonw} and Section \ref{sec:dSVDunknown}, respectively. The novel truncation technique is proposed in Section \ref{sec:dSVDknonw} alongside the first scenario. In Section \ref{sec:simulation}, two respective application examples for both proposed d-SVD algorithms are studied, i.e., the decentralized sensor localization, and the decentralized passive radar detection. We conclude our article in Section \ref{sec: conclusion}.

\textit{Notation:} The regular letter $a$, the boldface lowercase letter $\ba$, and the boldface uppercase letter $\bA$ denote a scalar, a column vector, and a matrix, respectively. If not otherwise specified, the $i$th row and the $i$th column of the matrix $\bA$ are denoted as $\bar{\ba}_i$ and $\ba_i$, respectively, and are treated as column vectors in equations. The calligraphic letter $\mathcal{A}$ denotes a set. The symbols $\mathbb{R}$ and $\mathbb{C}$ represent the real domain and the complex domain, respectively, and $(\cdot)^*$, $(\cdot)^\tT$ and $(\cdot)^\tH$ denote the conjugate, the transpose and the Hermitian operation, respectively. The operators $\odot$ and $\oslash$ denote the Hadamard, i.e., elementwise, product and division, respectively. Finally, the vector $\bone$ and $\bzero$ contain ones and zeros in all entries, respectively, $\bI$ is the identity matrix with a proper dimension, and $\be_i$ is the $i$th column of $\bI$.

%% file: pre.tex
\section{Preliminaries}\label{sec:pre}
In this work, we focus on implementing the SVD in a fully decentralized framework, where the matrix under consideration, denoted as $\bR\in\mathbb{C}^{N_1\times N_2}$, represents the data collected from various nodes distributed throughout the network. Let $N = \min\{N_1, N_2\}$ and the economy-size SVD of $\bR$ be
\begin{equation}\label{equ:svd}
	\bR = \bU\bSigma\bV^\tH,
\end{equation}
where $\bSigma = \diag{\sigma_1,\ldots,\sigma_N}\in\mathbb{R}^{N\times N}$ is the diagonal matrix that contains all singular values, and $\bU = [\bu_1,\ldots,\bu_{N}]\in\mathbb{C}^{N_1\times N}$ and $\bV = [\bv_1,\ldots,\bv_{N}]\in\mathbb{C}^{N_2\times N}$ are unitary matrices containing the left and right singular vectors, respectively. The network consisting of all nodes and their connections is characterized as the connected unweighted graph $\mathcal{G} = (\mathcal{V},\mathcal{E})$, where $\mathcal{V} = \{1,\ldots,N\}$ is the set of nodes and $\mathcal{E}\subseteq\mathcal{V}\times\mathcal{V}$ is the set of edges, i.e., the connections between nodes.

Based on the availability of $\bR$, we consider two scenarios of the d-SVD. First, the entries of $\bR$ are row-wisely available to the nodes in $\mathcal{G}$, i.e., the $i$th node has access to one row\footnote{The $i$th node may have access to multiple rows, which is a natural extension of our proposed algorithm, e.g., by substituting the consensus operation among available rows with local computation. Throughout the article, for simplicity of presentation, we assume that the knowledge in each node is limited to one row of $\bR$ or one element of $\bx(t)$ and $\by(t)$, respectively.} of $\bR$. Second, the matrix $\bR$ is not explicitly available, but is implicitly formed as a sum of outer products $\sum\bx(t)\by(t)^\tH$ of two different series of measurements $\bx(t)\in\mathbb{C}^{N}$ and $\by(t)\in\mathbb{C}^{N}$, and each node has access to one element of $\bx(t)$ and $\by(t)$, respectively, corresponding to its node index.

In our proposed algorithms for both scenarios of d-SVD, the in-network communication is carried out by any decentralized consensus scheme that computes the average (or summation) of graph signals, such as the Average Consensus (AC) algorithm \cite{xiaoFastlinear2004}, the Push-Sum (PS) algorithm \cite{kempeGossipbasedcomputation2003}, the finite-time Average Consensus (ftAC) algorithm \cite{kibangouFinitetimeaverage2011,sandryhailaFinitetimedistributed2014}, and the Privacy-Preserving Average Consensus (PPAC) algorithm \cite{liPrivacypreservingdistributed2020,liPrivacypreservingdistributed2022}. While the AC algorithm benefits from simple implementation, the PS algorithm is also applicable to directed graphs. Moreover, the ftAC algorithm guarantees a convergence in finite iterations, and the PPAC algorithm guarantees the privacy of local data if the privacy concern is critical. Throughout the article, we denote $\ps{\bx}$ and $\nc{k}{\bX,\bY}$ as network consensus operations that can be realized by any aforementioned consensus algorithms. Specifically, the operation $\ps{\bx}$ takes a vector argument $\bx$ with entry $x_i$ available in the $i$th node, and computes the summation $\sum_{i=1}^{N} x_i$ in each node using only local interaction. Similarly, the operation $\nc{k}{\bX,\bY}$ handles two matrix (or vector) arguments, $\bX$ and $\bY$, where the $i$th node has access to the rows $\bar{\bx}_i$ and $\bar{\by}_i$. This operation results in a matrix (or vector) available to all nodes that contains the inner product $\bX^\tH\bY$, where the subscript $k$ indicates the number of required parallel consensus instances.

\subsection{ EVD of the rank-one modification of a Hermitian matrix}\label{subsec:ra}
The SVD of the matrix $\bR$ is associated with the eigensystems of the symmetric matrices $\bR\bR^\tH$ and $\bR^\tH\bR$ \cite{golubMatrixcomputations2013}. Specifically, the left and right singular vectors of $\bR$ can be obtained by the EVD of $\bR\bR^\tH$ and $\bR^\tH\bR$, respectively. In this way, the problem of the SVD of a non-Hermitian matrix turns into the EVD of Hermitian matrices. Moreover, it has been shown in our previous work \cite{fanDecentralizedeigendecomposition2021,fanDecentralizedeigendecomposition2023} that the EVD of Hermitian matrices can be carried out in a fully decentralized manner with the help of the efficient RA approach. As will be shown later in Section \ref{sec:dSVDknonw} and Section \ref{sec:dSVDunknown}, both considered d-SVD scenarios can be transformed into the problem of a rank-one modified diagonal matrix, where the RA approach can be applied.

In the RA approach, we consider the EVD of a Hermitian matrix $\bR(t)\in\mathbb{C}^{N\times N}$ that is represented by the following rank-one modification of a Hermitian matrix
\begin{equation}\label{equ:rank1mod}
	\bR(t) = \bR(t-1) + \rho(t)\bx(t)\bx(t)^\tH,\quad t = 1,2,\ldots,
\end{equation}
where $t$ is the iteration index, $\bR(t-1)$ is Hermitian with $\bR(0) = \bzero$, $\rho(t)\in\mathbb{R}\setminus\{0\}$, and the $i$th entry of $\bx(t)$ is $x_i(t)$ that is only available to the $i$th node. Moreover, it is assumed that the eigenvalues $\boldsymbol{\Lambda}(t-1) = \diag{ \lambda_1(t-1),\ldots,\lambda_N(t-1)}$ and the corresponding eigenvectors $\bU(t-1) = [\bu_1(t-1),\ldots,\bu_N(t-1)]\in\mathbb{C}^{N\times N}$ of $\bR(t-1)$ are known and that the eigenvalues are distinct and sorted in a descending order as $\lambda_1(t-1)>\cdots>\lambda_N(t-1)$.

Multiplying both sides of \eqref{equ:rank1mod} with $\bU(t-1)^\tH$ and $\bU(t-1)$ from the left and the right, respectively, leads to
\begin{equation}\label{equ:rank1modified}
	\begin{aligned}
		\bU(t-1)^\tH \bR(t) \bU(t-1) = \boldsymbol{\Lambda}(t-1) + \rho(t) \bz(t)\bz(t)^\tH,
	\end{aligned}
\end{equation}
where
\begin{equation}\label{equ:zupdate}
	\bz(t)=\ 
	[z_{1}(t), \ldots, z_{N}(t)]^\mathsf{T} = 
	\bU(t-1)^\tH\bx(t).
\end{equation}
The expression on the right-hand side of (\ref{equ:rank1modified}) represents a rank-one modification of a diagonal matrix, whose EVD can be efficiently obtained via the RA approach \cite{liSolvingsecular1994,trinh-hoangPartialrelaxation2018}.

%

Denote the diagonal eigenvalue matrix and the corresponding eigenvector matrix of the right-hand side of \eqref{equ:rank1modified} as $\bar{\boldsymbol{\Lambda}}(t-1)=\diag{\bar{\lambda}_1(t-1),\ldots,\bar{\lambda}_N(t-1)}$ and $\bW(t-1) = [\bw_1(t-1),\ldots,\bw_N(t-1)]$, respectively, they are related as
\begin{equation}
	\label{equ:eigModMat}
		\medmath{\bW(t-1)^\tH\underbrace{(\boldsymbol{\Lambda}(t-1)+\rho(t)\bz(t)\bz(t)^\tH)}_{\bU(t-1)^\tH\bR(t)\bU(t-1)}\bW(t-1) = \bar{\boldsymbol{\Lambda}}(t-1).}
\end{equation}
Furthermore, since from the EVD of $\bR(t)$ we have
\begin{equation}
	\bU(t)^\tH\bR(t)\bU(t)=\boldsymbol{\Lambda}(t),
\end{equation}
it is observed that $\bR(t)$ in \eqref{equ:rank1mod} shares the same eigenvalues with the right-hand of \eqref{equ:rank1modified}, i.e.,
\begin{equation}\label{equ:eigValUpdate}
	\boldsymbol{\Lambda}(t) = \bar{\boldsymbol{\Lambda}}(t-1),
\end{equation}
and the corresponding eigenvectors are
\begin{equation}\label{equ:eigVecUpdate}
	\bU(t) = \bU(t-1)\bW(t-1).
\end{equation}

\subsection{Decentralized rational function approximation based eigenvalue decomposition (d-raEVD)}\label{subsec:oded}
In the aforementioned RA approach, four scalar parameters need to be computed in four non-linear equations (details cf. Appendix \ref{sec:appendixRA}) to obtain one eigenvalue of the right-hand side of \eqref{equ:rank1modified}. Moreover, closed-form expressions of these parameters only involve elementary arithmetic operations on scalar variables and are provided in \cite[Appendix G]{trinh-hoangPartialrelaxation2020}. Hence, each node in the network can feasibly implement and utilize the lightweight RA approach for the rank-one modified diagonal matrix\footnote{The RA approach is provided in the LAPACK \cite{andersonLAPACKusers1999} subroutines \texttt{dlaed4()} to compute the eigenvalues, and \texttt{dlaed9()}, which calls \texttt{dlaed4()}, to compute both the eigenvalues and the corresponding eigenvectors. In this work, we assume that each node in the network is facilitated with the \texttt{dlaed9()} subroutine, and uses it to compute the EVD of the rank-one modified diagonal matrix, i.e., the right-hand side of \eqref{equ:rank1modified}.}, e.g., the right-hand side of \eqref{equ:rank1modified}.

Nonetheless, in a fully decentralized setup, where no central unit collects all measurements $\bx(t)$, the RA approach is not applicable, as the rank-one update $\bz(t)$ in \eqref{equ:rank1modified} is not available to all nodes. If the data is available distributively and each node needs to broadcast its own data to all other nodes (or to the central unit), the $i$th node, $\forall i \in \mathcal{V}$, needs to store the instances, i.e., local copies, of the current and the past entire eigenvector matrices, denoted as $\bU_i(t)$ and $\bU_i(t-1)$, respectively, along with the local instances of the rank-one update, denoted as $\bz_i(t)$, at least one column of the auxiliary matrix $\bW_{i}(t-1)$, denoted as $\bw_{i,k}(t-1)\footnote{The indices $k$ and $t$ for auxiliary variables $\bz_i(t)$ and $\bw_{i,k}(t-1)$ are kept for the consistent presentation of the algorithm. In practical implementations, the storage for each auxiliary variable can be reused. Instead of storing the entire matrix $\bW(t)$, for example, every time a new eigenvector in $\bW(t)$ is computed via the RA approach, the update of the row of $\bU(t)$, i.e., the multiplication with $\bW(t)$, can be directly applied as indicated in \eqref{equ:eigVecUpdate}.}, k = 1,\ldots,N$, as well as the current and the past diagonal eigenvalue matrices, denoted as $\boldsymbol{\Lambda}_i(t)$ and $\boldsymbol{\Lambda}_i(t-1)$, respectively. This results in a total storage requirement of $2N^2 + 4N$ real floating point values in each node, which increases quadratically with the network size $N$. Hence, neither the centralized scheme nor the broadcasting scheme is scalable.

To reduce the storage cost in each node, by applying the aforementioned consensus algorithms to \eqref{equ:zupdate}, a fully decentralized rational function approximation based eigenvalue decomposition (d-raEVD) algorithm is proposed in our previous work \cite{fanDecentralizedeigendecomposition2021,fanDecentralizedeigendecomposition2023}, where we assume that the $i$th node stores only the $i$th row of $\bU(t-1)$ and $\bU(t)$, respectively, i.e., $\bar{\bu}_i(t)$ and $\bar{\bu}_i(t-1)$
, instead of the entire matrices. Thus, the storage cost in each node in the d-raEVD algorithm reduces to $6N$ real floating point values. For a detailed discussion regarding the storage cost cf. \cite{fanDecentralizedeigendecomposition2023}.

Following \eqref{equ:zupdate}, the $k$th entry of the rank-one update $\bz_i(t) = [z_{i,1}(t),z_{i,2}(t),\ldots,z_{i,N}(t)]^\tT$ is distributedly obtained via
\begin{equation}\label{equ:zdisupdate}
	z_{i,k}(t) = {\bu}_k(t-1)^\tH\bx(t) = \ps{{\bu}_k(t-1)\odot\bx(t)},
\end{equation}
and $\bz_i(t) = \nc{N}{\bU(t-1),\bx(t)}$. 
According to (\ref{equ:eigVecUpdate}), the entries of $\bar{\bu}_i(t)$ are locally updated by 
\begin{equation}\label{equ:eigVecUpdateLocal}
	\bar{u}_{i,k}(t) = \bar{\bu}_{i}(t-1)^\tH\bw_{i,k}(t-1),\quad k = 1, \ldots, N. 
\end{equation}

By initializing $\bLambda_i(0) = \bzero$ and $\bar{\bu}_i(0) = \be_i,$ $\forall i \in\mathcal{V}$, the d-raEVD algorithm distributedly computes the EVD of any matrices with the form of rank-one modified Hermitian matrix as shown in \eqref{equ:rank1mod}. 

\subsection{Remark on the implementation of d-raEVD}

For the practical implementation of the d-raEVD algorithm, two key issues must be addressed properly: (i) the multiplicity of the eigenvalues and (ii) the consistency of the solutions from the RA approach across nodes.

First, in the RA approach it is assumed that the eigenvalues of the symmetric matrix $\bR(t-1)$ are distinct. Nevertheless, this assumption may not always hold in practice. This is for example the case at the initial stage where all eigenvalues are initialized as zeros. To resolve this issue, following the discussion in \cite{bunchRankonemodification1978}, a deflation technique that utilizes the Householder transformation introduced in \cite{kuo-liangchungComplexHouseholder1997} has been proposed in our previous work \cite{fanDecentralizedeigendecomposition2023}. Particularly, such a deflation technique transforms the situation of singular values with multiplicities into a smaller dimension for the RA approach, which in fact simplifies the local computation in each node.

Second, in the RA approach it is required that the eigenvalues in $\bLambda(t-1)$ are descendingly ordered. Hence, at the end of the $t$th iteration, the resulted eigenvalues need to be sorted in a descending order, and the eigenvectors are sorted accordingly. In practice, due to machine precision limitations and the nature of decentralized consensus algorithms, each node may obtain slightly different local copies of $\bar{\bLambda}(t-1)$. For the special case when non-zero eigenvalues in $\bar{\bLambda}(t-1)$ have multiplicities, the local iterates of these eigenvalues may have minor mismatch, and hence, the sorting order of them may differ in different nodes. This leads to a local auxiliary matrix $\bW(t-1)$ in some nodes with swapped columns and an overall eigenvector matrix $\bU(t)$ updated by \eqref{equ:eigVecUpdate} with swapped entries. Thus, the final result of the d-raEVD may not be correct. Different methods can resolve this issue, and a local approach that does not require any extra communication cost is provided in Appendix \ref{sec:appendix1}. 

%% file: dSVDknown.tex
\section{D-SVD with Distributedly Partitioned Measurements}\label{sec:dSVDknonw}
In a decentralized network, the available entries of a non-symmetric or non-Hermitian matrix $\bR\in\mathbb{C}^{N\times T}$ are the data collected in each node, where $T$ is the sample size, and samples in each node are represented by the corresponding row of $\bR$. 
As mentioned in Section \ref{subsec:ra}, the left and right singular vectors of $\bR$ can be obtained via the EVD of
\begin{subequations}\label{equ:XUV}
	\begin{align}
		\bM_\text{U} =& \bR\bR^\tH =  \medop\sum_{t=1}^{T}\br(t)\br(t)^\tH = \bU_\text{U}\bLambda_\text{U}\bU_\text{U}^\tH,\label{subequ:XU}\\	
		\bM_\text{V} =& \bR^\tH\bR =  \medop\sum_{n=1}^{N}\bar{\br}(n)\bar{\br}(n)^\tH = \bV_\text{V}\bLambda_\text{V}\bV_\text{V}^\tH\label{subequ:XV},
	\end{align}
\end{subequations}
respectively, 
where $\br(t)$ and $\bar{\br}(n)$ are the $t$th column and $n$th row of $\bR$, respectively, and $\{\bU_\text{U}, \bLambda_\text{U}\}$ and $\{\bV_\text{V}, \bLambda_\text{V}\}$ are the eigenpairs of $\bM_\text{U}$ and $\bM_\text{V}$, respectively. To obtain the economy-size SVD of $\bR$ in a proper dimension, we define the extraction operator $\pi_{a,b}(\cdot)$ that selects the upper-left block with the size $a\times b$ in a matrix. Hence, the economy-size SVD of $\bR$ shown in \eqref{equ:svd} is extracted by
\begin{subnumcases}{\label{equ:RUVSigma}}
		\bU = \bU_\text{U},\label{subequ:RUVSigmaU}\\
		\bV = \pi_{T,N}(\bV_\text{V}),\label{subequ:RUVSigmaV}\\
		\bSigma = \bLambda_\text{U}^\frac{1}{2} = \pi_{N,N}(\bLambda_\text{V})^\frac{1}{2}.\label{subequ:RUVSigmaSigma}
\end{subnumcases}

Nevertheless, the d-raEVD approach introduced in Section \ref{subsec:oded} cannot be directly applied due to two following difficulties. First, although the EVD of $\bM_\text{U}$ can be performed by the d-raEVD approach, the same approach cannot be applied to compute the EVD of $\bM_\text{V}$. This is due to the fact that $\bar{\br}(n), n = 1,\ldots,N$, are the rows of $\bR$ where each row is available to one node, and cannot be treated as graph signals distributed over the network. Second, independently computing the left and right singular vectors via \eqref{subequ:XU} and \eqref{subequ:XV}, respectively, may not result in the correct singular vector pairs, since singular vectors corresponding to singular values with multiplicity are not unique. Specifically, the singular vectors are unique up to rotation or reflection \cite{huUniquenesssingular1997}, i.e., right multiplication with a unitary matrix $\bTheta$ to the left and to the right singular vector matrices, respectively, i.e., $\widetilde{\bU} = \bU\bTheta$ and $\widetilde{\bV} = \bV\bTheta$, maintains the unitary property of respective singular vector matrices, i.e., $\widetilde{\bU}^\tH\widetilde{\bU}=\bI$ and $\widetilde{\bV}^\tH\widetilde{\bV}=\bI$, and with $\widetilde{\bU}\bSigma\widetilde{\bV}^\tH = \bR$.
%
	
To overcome the two aforementioned difficulties, we introduce the following approach to find the associated left and right singular vectors of $\bR$. First, the d-raEVD approach is applied to $\bM_\text{U}$, and the singular values in $\bSigma$ and the left singular vector matrix $\bU$ are updated by \eqref{subequ:RUVSigmaU} and \eqref{subequ:RUVSigmaSigma}, respectively. Second, instead of computing the right singular vectors independently, the right singular vector associated with the $n$th positive singular value, i.e., $\sigma_n>0, \forall n = 1,\ldots,N,$ is distributedly computed as
\begin{equation}\label{equ:XUvn}
	\bv_n = \sigma_n^{-1}\nc{T}{\bR,\bu_n}, \quad n = 1,\ldots,N,
\end{equation}
and is available to all nodes. In total, $NT$ consensus instances and $T^2$ real floating point values in each node are required to compute and store $\bV$. 
Since $\bV$ is updated only once at the end of the d-SVD algorithm and is not involved in the iterative update of $\bSigma$ and $\bU$, the right singular vectors associated with zero singular values can be chosen as any vectors that maintain the orthogonality of $\bV$, e.g., by the distributed modified Gram-Schmidt (dmGS) orthogonalization algorithm \cite{strakovaDistributedQR2012}. It is worth mentioning that the necessity of computing $\bV$ depends on the specific application. Moreover, although each node results in full knowledge of $\bV$ after \eqref{equ:XUvn}, depending on practical scenarios, the storage cost in each node can be reduced, either by only maintaining one row or one column of $\bV$, or by directly applying columns of $\bV$ for further operation, e.g., multiplication (cf. Section \ref{subsec:loc}), once a column is obtained. We denote this proposed d-raEVD based SVD algorithm with distributedly partitioned measurements as d-raSVD1.


\subsection{Alternative: the decentralized Power Method (d-PM)}\label{subsec:dPM} The EVD of $\bM_\text{U}$ can also be obtained distributedly via the d-PM, and we denote the d-PM based d-SVD algorithm as d-pmSVD1. In the d-pmSVD1 approach, the storage assumption is the same as in the d-raSVD1 approach, i.e., each node maintains all eigenvalues and one row of the eigenvector matrix $\bU$, and the $n$th normalized eigenvector $\bu_n(p)$ of $\bM_\text{U}$ in \eqref{subequ:XU} is \cite{scaglioneDecentralizedestimation2008}
\begin{equation}\label{equ:dPMun}
	\bu_n(p) = \tilde{\bu}_n(p)/\norm{\tilde{\bu}_n(p)}_2, \quad n = 1,\ldots, N,
\end{equation}
where $p = 1,\ldots,P,$ is the iteration index of the PM and 
\begin{equation}\label{equ:dPMupdate}
	\begin{cases}
		\bu_n(p) = {\bu^\prime_n(p-1) - \displaystyle\sum_{q=0}^{n-1} \bu_q(p)\left(\bu_q(p)^\tH\bu_n^\prime(p-1)\right)},\\
		\bu_n^\prime(p-1) = {\displaystyle\sum_{n_2=1}^{N_2}\br(n_2)\left(\br(n_2)^\tH\bu_n(p-1)\right)}.
	\end{cases}
\end{equation}
In \eqref{equ:dPMupdate}, $\bu_0(p)=\bzero$, $\tilde{\bu}_n(0)$ is a random initial vector distributed across nodes. Specifically, the norm in \eqref{equ:dPMun} and the vector inner products in \eqref{equ:dPMupdate} are distributedly computed via $\ps{\bu_n(p)\odot\bu_n(p)}^\frac{1}{2}$, $\ps{\bu_q\odot\bu_n^\prime(p-1)}$, and $\ps{\br(t)\odot\bu_n(p-1)}$, respectively. 
Moreover, the associated eigenvalues are computed distributedly as
\begin{equation}
	\lambda_{\text{U},n} = \sigma_n^2 = \bu_n(P)^\tH\bu_n^\prime(P) = \ps{\bu_n(P)\odot\bu_n^\prime(P)},
\end{equation}
for $n = 1,\ldots, N$, and the right singular vectors $\bv_n$ can be obtained via \eqref{equ:XUvn}. Since the PM requires that the eigenvalues of $\bM_\text{U}$ are distinct, for the case with eigenvalue multiplicities, the d-pmSVD1 suffers from slow convergence speed, and thus, requires high communication cost.

\subsection{Communication cost and storage requirement of d-raSVD1 and d-pmSVD1}\label{subsec:dSVDknownCost}
\paragraph{Communication cost} We define the unit of the communication cost as one consensus instance that achieves one scalar summation\footnote{The consensus algorithms mentioned in Section \ref{sec:pre} are iterative and require multiple communications until convergence. Specifically, the ftAC algorithm converges in finite number of communications, i.e., in the number of the non-zero distinct eigenvalues of the graph Laplacian associated with the network.}. The communication cost of the d-raSVD1 algorithm associated with the update of the left singular vectors is $NT$ consensus instances that arises with $\bM_\text{U}$ in \eqref{subequ:XU}. Combining the number of consensus operation to compute the right singular vectors in \eqref{equ:XUvn}, i.e., $NT$ consensus instances, the proposed d-raSVD1 algorithm requires in total $2NT$ consensus instances. Since in the d-PM the subspace associated with the dominant eigenvector needs to be subtracted from the signal space to compute the second dominant eigenvector, and subsequently for the remaining eigenvectors, the total communication cost of d-PM is $N(TP+T+2) + PN(N-1)/2 + NT$ consensus instances \cite{fanDecentralizedeigendecomposition2023}, which is already much larger than that of our proposed d-raSVD1 algorithm.

\paragraph{Storage requirement} Assuming that the right singular vectors are not required, the storage requirement of the proposed d-raSVD1 algorithm is the same as that of the d-raEVD discussed in Section \ref{subsec:oded}, i.e., $6N$ real floating point values. In contrast, since the d-PM does not involve the auxiliary variable $\bW(t)$ but requires the local knowledge of all entries of the corresponding row of $\bR$ in each node, the associated storage requirement is $2N+T$ real floating point values.

\subsection{Truncate d-raEVD}\label{subsec:d-TraSVD}
Benefiting from the rank-one modification expression \eqref{equ:rank1mod}, the RA-based d-raEVD can be performed in an online manner each time a new sample is obtained. Nevertheless, all eigenvalues and associated eigenvectors are required to perform one update of d-raEVD based on a new sample, which is associated with significant undesired overhead for large-scale network applications where only the dominant eigenvalues are of interest, such as in PCA applications. In contrast, such issue is not encountered in the PM as the singular values and vectors are sequentially estimated (which, however, leads to slow convergence in the case where eigenvalues have multiplicities). For example, if only the largest eigenvalue is desired, $NT$ consensus instances are required to perform the complete d-raEVD algorithm while only $TP + T + 2$ consensus instances are sufficient for the d-PM algorithm. To overcome this drawback and noticing that all singular values are non-negative, we propose here a novel and non-trivial truncation approach, termed as d-TraEVD, that truncates the signal space of the desired signal to reduce the overall communication cost as well as the local computational overhead in each node.

Specifically, instead of the complete signal space, we only track the first $d \ll N$ largest eigenvalues and the associated eigenvectors that contain the principal eigenspace. Nevertheless, if the signal space is simply truncated to the subspace associated with the $d$ largest eigenvalues, only the signal components of $\bx(t)$ that correspond to the truncated subspace can contribute to the next update in \eqref{equ:rank1modified}, since the rank-one update $\bz(t)$ in \eqref{equ:zupdate} can be interpreted as the projection of $\bx(t)$ onto the truncated subspace. Consequently, the update $\bU(t)$ will lie in the same range space as $\bU(t-1)$ irrespectively the new datum $\bx(t)$. To allow the algorithm to escape from the range space of $\bU(t-1)$, we augment $\bU(t-1)$ by a so-called representative eigenvector, that is orthonormal to the truncated eigenspace of $\bU(t-1)$ to represent its complement subspace. Mathematically, in the $t$th iteration of the d-TraEVD approach, the new sample vector $\bx(t)$ is projected onto the truncated subspace, and the representative eigenvector $\hat{\bp}(t)$ is
\begin{equation}\label{equ:repreVec}
	\hat{\bp}(t) = \frac{\bx(t) - \widehat{\bU}(t-1)\widehat{\bU}(t-1)^\tH\bx(t)}{\norm{\bx(t) - \widehat{\bU}(t-1)\widehat{\bU}(t-1)^\tH\bx(t)}},
\end{equation}
where $\widehat{\bU}(t-1)\in\mathbb{C}^{N\times d}$ is the eigenvector matrix associated with the truncated subspace in the $(t-1)$th iteration. Consequently, the effective eigenspace applied in \eqref{equ:zupdate} is
\begin{equation}\label{equ:repreMat}
	\bU(t-1) = [\widehat{\bU}(t-1),\hat{\bp}(t)]\in\mathbb{C}^{N\times (d+1)}.
\end{equation}
Consequently, applying \eqref{equ:repreMat} in \eqref{equ:zupdate}, we obtain the effective rank-one update $\hat{\bz}(t) = [\widehat{\bU}(t-1),\hat{\bp}(t)]^\tH\bx(t)\in\mathbb{C}^{d+1}$ with the last entry
\begin{equation}\label{equ:effectRank1}
	\begin{aligned}
		&\hat{z}_{d+1}(t) = \hat{\bp}(t)^\tH\bx(t) = \left(\left\|\bx(t)\right\|^2 - \left\|\widehat{\bU}(t-1)^\tH\bx(t)\right\|^2\right)^\frac{1}{2},\\
		&=\left(\ps{\bx(t)\odot\bx(t)}^2 - \left\|\nc{d}{\widehat{\bU}(t-1),\bx(t)}\right\|^2 \right)^\frac{1}{2},
	\end{aligned}
\end{equation}
where $\nc{d}{\widehat{\bU}(t-1),\bx(t)}$ is the vector consisting the first $d$ entries of $\hat{\bz}(t)$ and only the norm of $\bx(t)$ requires one extra consensus instance.
%

\emph{Communication cost:} The required number of consensus instances by one rank-one update is reduced from $N$ in \eqref{equ:zupdate} to $d+1$ in \eqref{equ:effectRank1}, which includes the consensus instances required by the computation of the products $\widehat{\bU}(t-1)^\tH\bx(t)$ and the norm of $\bx(t)$. Moreover, the communication cost associated with the first $d$ right singular vectors in \eqref{equ:XUvn} is reduced from $NT$ to $dT$ consensus instances. Thus, the total communication cost of the d-raSVD1 is reduced from $2NT$ to $(2d+1)T$ consensus instances.

\emph{Storage requirement:} By truncating the d-raEVD to the first $d$ eigenvalues, not only the communication cost, but also the associated storage requirement is reduced. Specifically, the rational function approximation approach in each node operates on a lower dimension, i.e., $d+1$, instead of the dimension of the complete signal space, i.e., $N$. As a result, the required number of real floating point values is reduced from $6N$ in the complete d-raEVD to $6(d+1)$ in the d-TraEVD.

\emph{Numerical performance:}
\begin{figure}[tb]
	\centering
	\input{simulations/truncatedEVD_new}
	\caption{Error performance of the d-TraEVD with different $\delta$, where $N=100, T = 500$. Results are averaged over $200$ random realizations of $\bR$, and the relative error is examined only on the principal singular values.}
	\label{fig:truncatedEVD}
\end{figure}
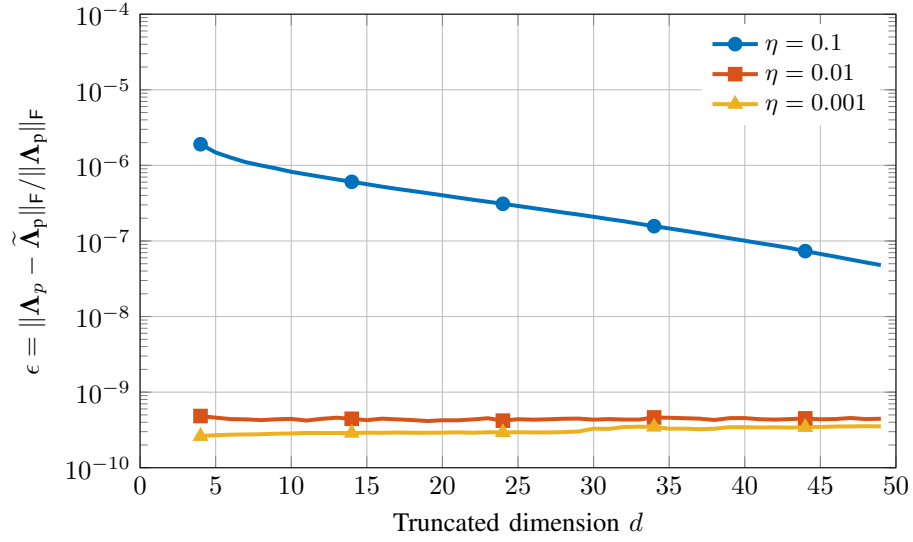
An Erd\H os R\'enyi network consists of $N = 100$ nodes is considered, where each node is initialized with $10$ neighbors and each connection is rewired with probability $0.1$. The entries of the matrix of interest $\bR \in \mathbb{R}^{N\times T}$, where $T = 500$, are drawn from the standard Gaussian distribution, and its principal subspace dimension is designed to be $4$ with the associated principal singular values $\sigma_1,\sigma_2,\sigma_3$ and $\sigma_4$. It is assumed that the largest singular value of the non-principal subspace, i.e., $\sigma_5$, is much smaller than the smallest singular value of the principal subspace, i.e., $\sigma_5 \leq \delta\sigma_4$ with $\delta\ll 1$ denoting the ratio. The consensus operation is carried out by the PS algorithm with $200$ iterations. The error performance 
is defined as $\epsilon = {\norm{\bSigma_\text{p}-\widetilde{\bSigma}_\text{p}}_\text{fro}}/{\norm{\bSigma_\text{p}}_\text{fro}}$, 
where each column of $\bSigma_\text{p}$ and of $\widetilde{\bSigma}_\text{p}$ is the vector containing the principal singular values and the local copies of the estimated principal singular values in each node, respectively. The simulation results illustrated in Fig.~\ref{fig:truncatedEVD} are averaged over $200$ random realizations of $\bR$. It is observed from Fig.~\ref{fig:truncatedEVD} that as the dimension $d$ of the truncated subspace increases, the accuracy of estimating the principal subspace via the d-TraEVD increases and approaches to that of the complete d-raEVD. Moreover, the error performance increases with increasing  gap $\delta$ between the principal and non-principal singular values.

%% file: simulations/truncatedEVD_new.tex
%
%
\definecolor{mycolor1}{rgb}{0.00000,0.44700,0.74100}%
\definecolor{mycolor2}{rgb}{0.85000,0.32500,0.09800}%
\definecolor{mycolor3}{rgb}{0.92900,0.69400,0.12500}%
\begin{tikzpicture}

\begin{axis}[%
width=10cm,
height=6.cm,
at={(0,0)},
scale only axis,
xmin=0,
xmax=50,
xminorticks=true,
xmajorgrids,
ymajorgrids,
ymode=log,
ymin=1e-10,
ymax=1e-4,
yminorticks=true,
xlabel=Truncated dimension $d$,
ylabel={$\epsilon = {\norm{\bLambda_p-\widetilde{\bLambda}_\text{p}}_\tF}/{\norm{\bLambda_\text{p}}_\tF}$},
mark size=2pt,
legend style={legend cell align=left, align=left, draw=white!15!black,font=\small,draw=none},
]

\addplot [color=mycolor1, line width=1.5pt,mark=*,mark repeat=10,mark options={solid}]
table[row sep=crcr]{%
	4	1.90392862491794e-06\\
	5	1.48124220840025e-06\\
	6	1.26651622024975e-06\\
	7	1.10103129229048e-06\\
	8	9.97934393499436e-07\\
	9	9.11711277907919e-07\\
	10	8.2201520756067e-07\\
	11	7.6110731044899e-07\\
	12	7.04528389135241e-07\\
	13	6.52921946954902e-07\\
	14	6.06079695845803e-07\\
	15	5.6378272045424e-07\\
	16	5.23102234259102e-07\\
	17	4.88572488423798e-07\\
	18	4.58249472896208e-07\\
	19	4.29636659012288e-07\\
	20	4.02207044974545e-07\\
	21	3.75695322932178e-07\\
	22	3.5145285875907e-07\\
	23	3.30542450329974e-07\\
	24	3.10141958930399e-07\\
	25	2.90426211112778e-07\\
	26	2.71971899618522e-07\\
	27	2.54460984989042e-07\\
	28	2.37722046471962e-07\\
	29	2.23413058234022e-07\\
	30	2.0897517988976e-07\\
	31	1.9459931183268e-07\\
	32	1.82401545929038e-07\\
	33	1.68762615936641e-07\\
	34	1.57074246856441e-07\\
	35	1.46404276450956e-07\\
	36	1.36429486929819e-07\\
	37	1.2668535955768e-07\\
	38	1.17285465987315e-07\\
	39	1.08527979687781e-07\\
	40	1.00793670160959e-07\\
	41	9.32848391201967e-08\\
	42	8.69062886091548e-08\\
	43	8.04551819672119e-08\\
	44	7.34270489674615e-08\\
	45	6.73692448633723e-08\\
	46	6.19725087558176e-08\\
	47	5.67687119219469e-08\\
	48	5.20641521081125e-08\\
	49	4.79492716553557e-08\\
};
\addlegendentry{$\eta = 0.1$}
%

\addplot [color=mycolor2, line width=1.5pt,mark=square*,mark repeat=10, mark options={solid}]
table[row sep=crcr]{%
	4	4.83354393039803e-10\\
	5	4.59955953342122e-10\\
	6	4.40070951340966e-10\\
	7	4.3627049633961e-10\\
	8	4.26579359029557e-10\\
	9	4.37285463662822e-10\\
	10	4.42736433198267e-10\\
	11	4.21174343317807e-10\\
	12	4.43336085555657e-10\\
	13	4.59056054367463e-10\\
	14	4.44073390222015e-10\\
	15	4.26095515661112e-10\\
	16	4.45661774143643e-10\\
	17	4.36362125434526e-10\\
	18	4.27694671210283e-10\\
	19	4.15046145402113e-10\\
	20	4.24674005014994e-10\\
	21	4.24046444632168e-10\\
	22	4.35266654577307e-10\\
	23	4.51918991067094e-10\\
	24	4.19508429959749e-10\\
	25	4.38438657183937e-10\\
	26	4.32461579902739e-10\\
	27	4.37249776976506e-10\\
	28	4.44846323653781e-10\\
	29	4.46654186024187e-10\\
	30	4.33070653553245e-10\\
	31	4.4047928169858e-10\\
	32	4.33637403706401e-10\\
	33	4.33275696822829e-10\\
	34	4.62852196263978e-10\\
	35	4.58169901431386e-10\\
	36	4.52195647697486e-10\\
	37	4.45917661170648e-10\\
	38	4.28183373792985e-10\\
	39	4.53539569689867e-10\\
	40	4.53709549030149e-10\\
	41	4.38049472436567e-10\\
	42	4.33256209370064e-10\\
	43	4.40357469975909e-10\\
	44	4.49717823576981e-10\\
	45	4.37352127947038e-10\\
	46	4.40515409069301e-10\\
	47	4.55429368169991e-10\\
	48	4.39988973471248e-10\\
	49	4.45038502510585e-10\\
};
\addlegendentry{$\eta=0.01$}


\addplot [color=mycolor3, line width=1.5pt,mark=triangle*,mark repeat=10, mark options={solid}]
table[row sep=crcr]{%
	4	2.60998947447155e-10\\
	5	2.69646253434016e-10\\
	6	2.73761396248155e-10\\
	7	2.75674383651959e-10\\
	8	2.77699700702879e-10\\
	9	2.81741418854385e-10\\
	10	2.84251245798834e-10\\
	11	2.88725107675907e-10\\
	12	2.8804944264706e-10\\
	13	2.88150503259786e-10\\
	14	2.87922133405767e-10\\
	15	2.90551115499668e-10\\
	16	2.89751767404106e-10\\
	17	2.92136790494617e-10\\
	18	2.90119676664288e-10\\
	19	2.89974919397942e-10\\
	20	2.91292900422389e-10\\
	21	2.94130624250926e-10\\
	22	2.90214778130206e-10\\
	23	2.94952351024584e-10\\
	24	2.93501139466734e-10\\
	25	2.94911338863715e-10\\
	26	2.93137119664416e-10\\
	27	2.92938449787242e-10\\
	28	2.96163230287924e-10\\
	29	3.0172287593297e-10\\
	30	3.28961342386381e-10\\
	31	3.26498161031979e-10\\
	32	3.46048060781621e-10\\
	33	3.48873994785434e-10\\
	34	3.4680915515199e-10\\
	35	3.28260606225733e-10\\
	36	3.28878437673235e-10\\
	37	3.23242185175747e-10\\
	38	3.28179093089091e-10\\
	39	3.44446140695021e-10\\
	40	3.43561233822437e-10\\
	41	3.40705302063648e-10\\
	42	3.42575804053465e-10\\
	43	3.39822425554918e-10\\
	44	3.4289825256381e-10\\
	45	3.46151086267513e-10\\
	46	3.50999376961823e-10\\
	47	3.5218999450738e-10\\
	48	3.54224682132802e-10\\
	49	3.52351421420171e-10\\
};
\addlegendentry{$\eta=0.001$}
\end{axis}

%
%
%
\end{tikzpicture}%

%% file: dSVDunknown.tex
\section{D-SVD of an Implicit Outer Product}\label{sec:dSVDunknown}
The matrix under consideration $\bR$ may not always be directly available in the nodes and is often only implicitly obtained, e.g., from two distinct measurement vectors. 
Mathematically, $\bR$ is expressed as
\begin{equation}\label{equ:Runsym}
	\bR = \bR(T) = {\sum_{t=1}^{T}}\bx(t)\by(t)^\tH = \bX\bY^\tH,
\end{equation}
where $\bx(t)\in\mathbb{C}^{N}$ and $\by(t)\in\mathbb{C}^{N}$ denote the first and the second measurement vector, respectively, and $\bX = [\bx(1),\ldots,\bx(T)]\in\mathbb{C}^{N\times T}$, $\bY = [\by(1),\ldots,\by(T)]\in\mathbb{C}^{N\times T}$. Note that each node in the network has only access to one corresponding element of $\bx(t)$ and $\by(t)$, respectively, or equivalently, one row of $\bX$ and $\bY$, respectively. Similar to \eqref{equ:RUVSigma}, the economy-size SVD of $\bR$ is related to the EVD of the auxiliary variables $\bM_\text{U}$ and $\bM_\text{V}$ defined in \eqref{equ:XUV} as
\begin{equation}\label{equ:R2UVSigma}
		\bU = \bU_\text{U},\
		\bV = \bV_\text{V},\text{ and }
		\bSigma = \bLambda_\text{U}^\frac{1}{2} = \bLambda_\text{V}^\frac{1}{2}.
\end{equation}

Although the matrix $\bR$ can be expressed as 
\begin{equation}\label{equ:Xrank1nonsymm}
	\bR(t) = \bR(t-1) + \bx(t)\by(t)^\tH,\quad t = 1, \ldots, T,
\end{equation}
where $\bR(0) = \bzero$, it differs from the case mentioned in Section \ref{sec:dSVDknonw}, i.e., a rank-one modified Hermitian matrix, as $\bR(t)$ is generally non-symmetric or non-Hermitian. Thus, the aforementioned d-raSVD1 algorithm is not applicable. Nevertheless, the auxiliary matrices $\bM_\text{U}$ and $\bM_\text{V}$ in \eqref{equ:XUV} can be reformulated in the form of consecutive rank-one modifications of a Hermitian matrix, and the SVD of $\bR$ can be carried out in a fully decentralized manner, which we denote as d-raSVD2. Throughout this section, it is assumed that the SVD of $\bR(t-1)$ is available distributively across nodes, i.e., each node has access to the complete set of the singular values in $\bSigma(t-1)$ and one row of the left and right singular vector matrices $\bU(t-1)$ and $\bV(t-1)$, respectively. Under this assumption, the proposed d-raSVD2 algorithm is an online algorithm that tracks the SVD of $\bR(t)$ once a new pair of samples/measurements of $\bx(t)$ and $\by(t)$ is obtained, since only the SVD of $\bR(t-1)$ is required to update the SVD of $\bR(t)$. 

\subsection{Update of the left singular vectors}\label{subsec:updateleft}
We start with the expression of $\bM_\text{U}(t)$ in \eqref{subequ:XU} and reformulated it as \cite{stangeEfficientupdate2008}
\begin{equation}\label{equ:MUrank2}
	\begin{aligned}
		\bM_\text{U}(t) &= \bM_\text{U}^\prime(t)+ \gamma_{2}{\bq}_{2}(t){\bq}_{2}(t)^\tH,
	\end{aligned}
\end{equation}
where
\begin{subequations}
	\begin{align}
		\bM^\prime_\text{U}(t)& = \bM_\text{U}(t-1) + \gamma_{1}{\bq}_{1}(t){\bq}_{1}(t)^\tH,\label{subequ:MUprime}\\
		[{\bq}_{1}(t), {\bq}_{2}(t)]& = {\bQ}_\beta(t)\bPhi,\label{subequ:xtildem}\\
		\bQ_\beta(t)& = [\bx(t),\tilde{\by}(t)],\label{subequ:Qbeta}\\
		\tilde{\by}(t) &= \bR(t-1)\by(t),\label{subequ:xm2tilde}
	\end{align}
\end{subequations}
and $\boldsymbol{\Gamma}=\diag{\gamma_{1},\gamma_{2}}$ and $\bPhi = [\bphi_{1},\bphi_{2}]$ are the respective eigenvalue and eigenvector matrix of $\bB = \begin{medsize}
	\begin{bmatrix}
	\beta & 1\\
	1 & 0
\end{bmatrix}\end{medsize}$ with $\beta = \norm{\by(t)}^2$. 
%
Substituting \eqref{equ:svd} in \eqref{subequ:xm2tilde} results in
\begin{equation}\label{equ:ybar}
	\tilde{\by}(t) = \bU(t-1)\bSigma(t-1)\bz_\beta(t),
\end{equation}
where
\begin{equation}\label{equ:zbeta}
	\bz_\beta(t) = \bV(t-1)^\tH\by(t) = \nc{N}{\bV(t-1),\by(t)}.
\end{equation}
Thus, the $i$th entry in $\tilde{\by}(t)$ is computed locally as $\tilde{y}_{i}(t) = \bar{\bu}(t-1)^\tT\bSigma(t-1)\bz_\beta(t)$, 
where $\bar{\bu}(t-1)$ is the $i$th row of $\bU(t-1)$ that is only available to the $i$th node.

Since $\beta = \ps{\by(t)\odot\by(t)}$ is available to all nodes, the computation of the EVD of $\bB$ is obtained locally in each node. Moreover, the associated computational cost is negligible as the dimension of $\bB$ is small, i.e., $2\times 2$. Hence, $\boldsymbol{\Gamma}$ and $\bPhi$ are available to all nodes. Moreover, each node has access to one row of ${\bQ}_\beta(t)$ corresponding to its node index, and the corresponding entries in ${\bq}_{1}(t)$ and ${\bq}_{2}(t)$ shown in \eqref{subequ:xtildem} are computed locally in each node. Hence, the expression of $\bM_\text{U}(t)$ in \eqref{equ:MUrank2} is a two consecutive rank-one modifications of a Hermitian matrix. Thus, the EVD of $\bM_\text{U}(t)$ can be obtained after two applications of the RA approach with respect to ${\bq}_{1}(t)$ and ${\bq}_{2}(t)$.

Specifically, the eigenvector matrix $\bU^\prime(t)$ and the diagonal eigenvalue matrix $\bLambda^\prime(t)$ of $\bM^\prime_\text{U}(t)$ are computed distributedly by applying the d-raEVD algorithm mentioned in Section \ref{subsec:oded} to \eqref{subequ:MUprime}, where the corresponding rank-one update is computed distributedly via $\nc{N}{\bU(t-1),\bq_{1}(t)}$. Denoting $\bW^\prime(t-1)$ as the eigenvector matrix associated with the rank-one modification of the diagonal matrix $\bLambda(t-1)$, which is accessible in each node, similar to \eqref{equ:eigVecUpdate}, the eigenvector matrix $\bU^\prime(t)$ is updated as
\begin{equation}\label{equ:Uprime}
	\bU^\prime(t) = \bU(t-1)\bW^\prime(t-1).
\end{equation}
As a result, each node has access to the complete set of eigenvalues in $\bLambda^\prime(t)$ and one row of $\bU^\prime(t)$ that is associated with its node index. 

Similarly, after decentralized computing the rank-one update associated with \eqref{equ:MUrank2}, i.e., $\nc{N}{\bU^\prime(t),\bq_2(t)}$, the eigenvalue matrix $\bLambda_\text{U}(t)$ of $\bM_\text{U}(t)$ in \eqref{equ:MUrank2} is obtained via the d-raEVD. The associated eigenvector matrix $\bU_\text{U}(t)$ is updated as
\begin{equation}\label{equ:UU}
	\bU_\text{U}(t) = \bU^\prime(t)\bW^{\prime\prime}(t-1),
\end{equation}
where $\bW^{\prime\prime}(t-1)$ is the eigenvector matrix associated with the rank-one modification of the matrix $\bLambda^\prime(t)$. Substituting the obtained matrices $\bLambda_\text{U}(t)$ and $\bU_\text{U}(t)$ in \eqref{equ:R2UVSigma} results in
\begin{equation}
	\bSigma(t) = \bLambda_\text{U}(t)^\frac{1}{2} \text{ and } \bU(t) = \bU_\text{U}(t).
\end{equation}
Thus, the updates of the left singular vector matrix $\bU(t)$ and the singular value matrix $\bSigma(t)$ of $\bR(t)$ are distributedly obtained.
%
%
%

\subsection{Update of the right singular vectors}\label{subsec:updateright}
The update of the right singular vectors can be obtained by applying the approach introduced in Section \ref{subsec:updateleft}. Nevertheless, since the SVD of a matrix is generally not unique \cite{huUniquenesssingular1997}, as mentioned in Section \ref{sec:dSVDknonw}, the unitary transformation ambiguity of the singular vectors arises from singular value multiplicities. To this end, we update the right singular vectors based on the updated left singular vectors as follows.

From \eqref{equ:svd} and \eqref{equ:Xrank1nonsymm}, we have
\begin{equation}\label{equ:Xrank1nonsymmexp}
	{\bU(t)\bSigma(t)\bV(t)^\tH = \bU(t-1)\bSigma(t-1)\bV(t-1)^\tH + \bx(t)\by(t)^\tH.}
\end{equation}
Multiplying $\bU(t)^\tH$ to the left and the right side of \eqref{equ:Xrank1nonsymmexp} and taking the Hermitian of both sides result in
\begin{equation}\label{equ:V+update}
	\Breve{\bV}(t) = \Breve{\bV}(t-1)\Breve{\bU}(t)^\tH + \by(t)\breve{\bx}(t)^\tH,
\end{equation}
where
\begin{subequations}
	\begin{align}
		\Breve{\bV}(t) &= \bV(t)\bSigma(t), \label{subequ:V+}\\
		\Breve{\bV}(t-1) &= \bV(t-1)\bSigma(t-1), \label{subequ:Vt-1+}\\
		\Breve{\bU}(t) &= \bU(t)^\tH\bU(t-1), \label{subequ:UhatOld}\\
		\breve{\bx}(t) &= \bU(t)^\tH\bx(t).\label{subequ:xhat}
	\end{align}
\end{subequations}
Moreover, substituting \eqref{equ:UU} and \eqref{equ:Uprime} in \eqref{subequ:UhatOld}, $\Breve{\bU}(t)$ can be equivalently updated by the product of two local intermediate auxiliary variables as
\begin{equation}\label{equ:Uhat}
	\Breve{\bU}(t) = \bW^{\prime\prime}(t-1)^\tH\bW^\prime(t-1)^\tH.
\end{equation}

Recall that following the $t$th update of the left singular vectors in Section \ref{subsec:updateleft}, each node has access to the complete set of matrices $\bSigma(t-1), \bSigma(t)$, $\bW^\prime(t-1)$, and $\bW^{\prime\prime}(t-1)$, one corresponding row of $\bU(t-1)$, $\bU(t)$, and $\bV(t-1)$, respectively, and one corresponding entry of $\bx(t)$ and $\by(t)$, respectively. Hence, while \eqref{equ:Uhat} is computed locally, \eqref{subequ:xhat} is obtained via $\nc{N}{\bU(t),\bx(t)}$. Moreover, the rows of $\Breve{\bV}(t-1)$ in each node are computed locally via \eqref{subequ:Vt-1+}. Consequently, the update of the rows of $\Breve{\bV}(t)$ is obtained locally via \eqref{equ:V+update}. From \eqref{subequ:V+}, the right singular vectors are updated as
\begin{equation}\label{equ:VV+}
	\bV(t) = \Breve{\bV}(t)\bSigma(t)^{-1}.
\end{equation}

\subsection{Communication cost and storage requirement of d-raSVD2 and d-pmSVD2}
\paragraph{Communication cost} In each iteration, the decentralized updates of $\beta$ in $\bB$, $\bz_\beta(t)$ in \eqref{equ:zbeta}, and $\breve{\bx}(t)$ in \eqref{subequ:xhat} require $1$, $N$, and $N$ consensus instances, respectively. Additionally, the consecutive rank-one modification expression in \eqref{equ:MUrank2} requires $2N$ consensus instances to compute the two rank-one updates. In total, $T(4N+1)$ consensus instances are required to perform the d-raSVD2 approach. Hence, the communication cost increases as the the number of nodes $N$ in the network increases, which, depending on the application, can be reduced by applying the truncation technique introduced in Section \ref{subsec:d-TraSVD}. We denote the truncated version of d-raSVD2 as d-TraSVD2($d$), and the required number of consensus instances is reduced to $4(d+1)T$.
\paragraph{Storage requirement} The right singular vectors are not necessarily stored during the update of the proposed d-raSVD2 algorithm. In fact, we can use the intermediate variable $\Breve{\bV}(t)$ to perform the consensus operation and to compute $\tilde{\by}(t)$ in \eqref{equ:ybar}, and then obtain $\bV(t)$ only in the final step of the d-raSVD2 algorithm via \eqref{equ:VV+}. In this case, apart from the storage requirement of the d-raEVD algorithm, only the scalar values $\gamma_{1},\gamma_{2}$, one corresponding entry of $\bq_{1}(t)$ and $\bq_{2}(t)$, respectively, and one corresponding row of $\Breve{\bV}(t-1)$ and $\Breve{\bV}(t)$, respectively, need to be stored in each node. Hence, the total storage requirement of the proposed d-raSVD2 algorithm is $8N+4$ real floating point values. If the truncation technique is applied, the storage requirement of the d-TraSVD2($d$) is further reduced to $8(d+1)+4$ real floating point values.
\begin{remark}
	A PM based d-SVD algorithm that is proposed in \cite{wuDecentralizedarray2019} computes the left and right singular vectors $\bu_1$ and $\bv_1$ associated with the largest singular value $\sigma_1$ iteratively as:
	\begin{equation}\label{equ:powerSVD}
		\begin{cases}
			\bu_1(p) &= \bX\hat{\bv}_1(p) + \alpha\bu_1(p-1),\ \hat{\bv}_1(p) = \bY^\tH\bv_1(p-1),\\
			\bv_1(p) &= \bY\hat{\bu}_1(p) + \alpha\bv_1(p-1),\ \hat{\bu}_1(p) = \bX^\tH\bu_1(p-1),
		\end{cases}
	\end{equation}
	where $\bu_1(0)$ and $\bv_1(0)$ are random vectors, $\hat{\bv}_1(p) =\nc{T}{\bY,\bv_1(p-1)}$ and $\hat{\bu}_1(p) = \nc{T}{\bX,\bu_1(p-1)}$, and $0<\alpha<1$. Accordingly, the largest singular value is obtained via $\sigma_1 = \hat{\bu}_1(P)^\tH\hat{\bv}_1(P)$. We denote \eqref{equ:powerSVD} as the d-pmSVD2 approach. Similar to the d-PM \cite{scaglioneDecentralizedestimation2008}, all measurements of $\bx(t)$ and $\by(t)$ are required to compute $\hat{\bu}_1(p)$ and $\hat{\bv}_1(p)$ in each iteration. Thus, unlike the online d-raSVD2 approach, the d-pmSVD2 approach is a batch approach that requires complete knowledge of measurements $\bX$ and $\bY$ in each iteration. In addition, the communication cost of the d-pmSVD2 approach associated with only the largest singular value is $P(2T+2)+2T$ consensus instances, which includes two normalization operations in each PM iteration.
\end{remark}

%% file: simulation.tex
\section{Application Examples and Simulation Results}\label{sec:simulation}
In this section, we examine the performance of our proposed d-SVD algorithms, and study application examples that rely on the d-SVD, i.e., the decentralized sensor localization and the decentralized passive radar detection.

\input{dLocalization}
\subsection{Decentralized passive radar detection}
In wireless communication systems, passive detection of potential airborne targets can be achieved by exploiting signals received from non-cooperative illuminators, such as existing base stations or TV towers with multiple transceivers. This approach holds significant interest in both civilian and military scenarios \cite{liuPerformancecrosscorrelation2015}. In distributed WSN systems, decentralized passive radar detection may be applied using the proposed d-raSVD2 algorithm without collecting measurements of all radars in a fusion center. A decentralized passive radar system is described by $L$ non-cooperative illuminators, e.g., $L$ broadcasting TV towers, and $N$ pairs of passive radars. These passive radars and their communication links are characterized by a graph $\mathcal{G}$, where each pair of passive radars is represented by a node in $\mathcal{G}$. Moreover, in each pair of passive radars, one radar serves as the reference channel that receives only signals directly from the non-cooperative illuminators, while the other one serves as the surveillance channel that receives both the signals obtained directly from the non-cooperative illuminators and the signals reflected from the targets. It is assumed that the illuminators transmit uncorrelated signals over a common bandwidth and the signals received at the reference channel and the surveillance channel are synchronized to deal with the delay and Doppler effect \cite{santamariaPassivedetection2017}. An example of such decentralized passive radar systems is illustrated in Fig.~\ref{fig:passiveRadar}, where the passive radars facing left represent the reference channel and those facing right represent the surveillance channel.
\begin{figure}[tb]
	\centering
	\input{simulations/passiveRadar1}
	\caption{An example of distributed passive radar system with $L=1$ non-cooperative illuminator.}
	\label{fig:passiveRadar}
\end{figure}

\paragraph{System model} Let $\bkappa(t)\in\mathbb{C}^L, t = 1,\ldots,T,$ denote the signal transmitted by the $L$ non-cooperative illuminators, the observed signals at the reference channel and the surveillance channel are $\br(t) = \bH_r\bkappa(t) + \bn_r(t)\in\mathbb{C}^{N}$ and $\bs(t) = \xi\bH_s\bkappa(t) + \bH_r\bkappa(t) + \bn_s(t)\in\mathbb{C}^{N}$, respectively, where $\xi\in\{0,1\}$, $\bH_r\in\mathbb{C}^{N\times L}$ and $\bH_s\in\mathbb{C}^{N\times L}$ are the effective channel matrices between the illuminators and the reference and the surveillance channel, respectively, and $\bn_r(t)\in\mathbb{C}^{N}$ and $\bn_s(t)\in\mathbb{C}^{N}$ represent the uncorrelated noise at the reference and at the surveillance channel, respectively. Note that each node in $\mathcal{G}$ has access to one corresponding element of $\br(t)$ and $\bs(t)$, respectively. 
Moreover, $\xi = 0$ under the null hypothesis $\mathcal{H}_0$, and $\xi = 1$ under the alternative $\mathcal{H}_1$, when the target is present.

\paragraph{Decentralized cross-correlation detector} The cross-correlation detector in the multi-antenna case is \cite{liuPerformancecrosscorrelation2015,santamariaPassivedetection2017}
\begin{equation}\label{equ:cross-corr}
	\left|\tr{\bR_{sr}^\tH\bR_{sr}}\right|\ \mathop{\gtrless}_{\mathcal{H}_0}^{\mathcal{H}_1}\ \eta,
\end{equation}
where $\bR_{sr}$ is the empirical cross-covariance matrix obtained by $\bR_{sr} = \frac{1}{T}\sum_{t=1}^{T}\bs(t)\br(t)^\tH$ and $\eta$ is an appropriate {thres-hold}. 
Denote $\sigma_n,$ $n = 1,\ldots,N,$ as the singular values of $\bR_{sr}$, \eqref{equ:cross-corr} is equivalent to
\begin{equation}
	{\sum_{n=1}^{N}}\sigma_n^2 \ \mathop{\gtrless}_{\mathcal{H}_0}^{\mathcal{H}_1}\ \eta.
\end{equation}
Since only the singular values are required to perform the cross-correlation detector, it is straightforward to apply the proposed d-raSVD2 algorithm in Section \ref{sec:dSVDunknown} to the two series of measurements, i.e., $\br(t)$ and $\bs(t)$.

\paragraph{Experimental results} In the simulations, a single non-cooperative illuminator is considered, i.e., $L=1$, with $\kappa(t)\sim\mathcal{CN}(0,1)$. The channel matrices $(\bH_r,\bH_s)$ and noise vectors $(\bn_r(t),\bn_s(t))$ are drawn from the standard Gaussian distribution. In addition, $(\bH_r,\bH_s)$ are scaled with respect to the Signal-to-Noise Ratio (SNR), where the SNR is fixed at $-10$ dB for both the reference and surveillance channels. The probability of detection $P_\text{d}$ and the probability of false alarm $P_\text{fa}$ are estimated by averaging $10^4$ independent channel realizations. As a result, the Receiver Operating Characteristic (ROC) curves of a small world example, i.e., $N =10, T=5$, are obtained as shown in Fig.~\ref{fig:passiveRadarResult}, where the decentralized power SVD, termed as d-pmSVD2 and the decentralized SVD using the d-TraEVD algorithm with truncation dimension $d = 1$, termed as d-TraSVD2(1), is implemented for comparisons. Moreover, the associated representative graph $\mathcal{G}$ is an Erd\H os R\'enyi network with the rewire probability $0.1$. The PS iteration number for all decentralized algorithms is $10N$. In the d-pmSVD2 algorithm, the power iteration is $P = 10$, $\alpha = 0.1$, and only the largest singular value is computed to reduce the communication cost.

\begin{figure}[ht]
	\begin{minipage}{.5\linewidth}
		\centering
		\input{simulations/passiveRadarSimulation_new}
		\caption{ROC curves for $L = 1, N = 10, T = 5,$  SNR $=-10$ dB, $P = 10, \alpha=0.1$.}
		\label{fig:passiveRadarResult}
	\end{minipage}
	\hfill
	\begin{minipage}{.5\linewidth}
		\centering
		\input{simulations/passiveRadarSimulation100_new}
		\caption{ROC curves for $L = 1, N = 100, T = 50,$  SNR $=-10$ dB, $P = 10, \alpha=0.1$.}
		\label{fig:passiveRadarResult100}
	\end{minipage}
\end{figure}

From Fig.~\ref{fig:passiveRadarResult} we observe that the d-raSVD2, the d-pmSVD2, and the d-TraSVD2($1$) algorithms achieve the theoretical bound (overlapped with each other) that is obtained via centralized computation. The d-pmSVD2 algorithm, which computes only the largest singular value, has an overall communication cost of $P(2T+2)+2T=130$ consensus instances. In contrast, the d-raSVD2 algorithm, which computes all singular values, has a higher communication cost of $T(4N+1)=205$ consensus instances. Moreover, the d-TraSVD2(1) requires only $4(d+1)T = 40$ consensus instances, which is $19.51\%$ and $30.77\%$ of the d-raSVD2 and of the d-pmSVD2, respectively, and achieves a similar ROC curve as the theoretical bound.

Similar simulation performance is observed in a larger network with $N =100$ nodes and with $T=50$ samples, as shown in Fig.~\ref{fig:passiveRadarResult100} while other simulation parameters remain the same. As the network size and number of samples increase, the overall communication cost of each algorithm increases. Specifically, while the communication cost of the d-pmSVD2 increases linearly (since only the largest singular value is computed) with $T$ and results in $1120$ consensus instances, the communication cost of the proposed d-raSVD2 approach is affected both by $N$ and $T$, and results in $20050$ consensus instances. Nevertheless, benefiting from the truncation technique, the communication cost of the d-TraSVD2($1$) is significantly reduced to $400$ consensus instances while achieving a similar ROC curve as the centralized solution. The overall communication cost comparison is summarized in Table~\ref{tab:comm}.

%

\begin{table}[ht]
	\centering
	\begin{talltblr}[
		caption = {Communication cost of d-SVD2 algorithms.},
		label = {tab:comm},
		note{*} = {\scriptsize Only the largest singular value is computed and the normlaizationed in each power iteration are considered.},
		]{hlines, vlines, colspec = {l*{3}c},
				cell{1}{1} = {c=2}{c},
				}
		$\{N,T\}$ & & $\{10,5\}$ &  $\{100,50\}$ \\
		d-pmSVD2\TblrNote{*} & $P(2T+2)+2T$& $130$ & $1120$\\
		d-raSVD2 & $T(4N+1)$ & $205$ & $20050$ \\
		d-TraSVD2($1$) & $4(d+1)T$& $40$ & $400$\\
	\end{talltblr}
\end{table}


%% file: dLocalization.tex
\subsection{Decentralized sensor localization}\label{subsec:loc}
Providing the physical localization of antennas or antenna arrays as well as the positions of the respective mobile users can be utilized to provide and improve many location-based services \cite{schillerLocationbasedservices2004,lemicLocalizationfeature2016}, and can be found in various study items in the 3rd Generation Partnership Project (3GPP) \cite{wenSurvey5G2019}. In this application example, we want to design a decentralized localization approach that is based only on local measurements of the mutual inter-sensor distances, where each antenna or antenna array is considered as one sensor. Different techniques are available to estimate the inter-sensor distances, such as the Angle-of-Arrival (AoA), the Time-of-Arrival (ToA), and the Received Signal Strength Indicator (RSSI) \cite{vuckovicLocalizationtechniques2023}, where Line-of-Sight (LoS) paths between sensors are preferred to obtain improved estimation accuracy. However, due to, e.g., the existence of obstacles, not all LoS paths are available in practice. In our decentralized localization application, only the mutual inter-sensor distances associated with LoS communication links are considered. Moreover, such mutual inter-sensor distances are only shared between both ends of the LoS communication links, and are not collected in a fusion center.

In this application, each sensor is characterized as a node in the network as well as in the representative graph $\mathcal{G}$, where undirected edges represent the LoS communication links. Denote the true coordinate of the $i$th node as $\bx_{i}\in\mathbb{R}^{h}$, where $h$ is the dimension of the node space. The complete coordinate matrix of the node network is $\bX = [\bx_{1},\ldots,\bx_{N}]\in\mathbb{R}^{h\times N}$. In addition, denote the symmetric Euclidean Distance Matrix (EDM) as $\bR \in\mathbb{R}^{N\times N}$ that contains squared distances between nodes in the network, where $r_{ij} = \norm{\bx_{i}-\bx_{j}}_2^2$, $i, j = 1,\ldots, N,$ is the squared distance between the $i$th and the $j$th node. 
Moreover, the EDM $\bR$ is related to $\bX$ as 
\begin{equation}\label{equ:edm}
	\bR = \bone\diag{\bX^\tT\bX}^\tT - 2\bX^\tT\bX + \diag{\bX^\tT\bX}\bone^\tT.
\end{equation}

If the entire matrix $\bR$ is available, the coordinates of all nodes can be estimated via the classical Multi-Dimensional Scaling (MDS) algorithm \cite{torgersonMultidimensionalscaling1965} up to rotation, reflection, and translation \cite{dokmanicEuclideandistance2015}. However, in practice, the available EDM $\widehat{\bR} = \bR\odot\bA$, where $\bA$ is the adjacency matrix of $\mathcal{G}$, may not be complete due to, e.g., obstacles between nodes. Nonetheless, noticing from \eqref{equ:edm} that the rank of $\bR$ is at most $h+2$ \cite{dokmanicEuclideandistance2015}, low-rank matrix completion techniques, such as the Singular Value Thresholding (SVT) algorithm \cite{caiSingularvalue2010} can be applied to estimate the complete EDM. We propose to use the d-raSVD1 algorithm to perform decentralized sensor localization using a communication efficient fully decentralized implementation of the SVT algorithm.

\paragraph{Decentralized Singular Value Thresholding (d-SVT)} The SVT algorithm consists in the following updates
\begin{subnumcases}{\label{equ:svt}}
	\widetilde{\bR}(t) = \mathcal{D}_\tau\left(\bR(t-1)\right),\label{subequ:shrink}\\
	\bR(t) = \bR(t-1) + \mu\left(\widehat{\bR} - \widetilde{\bR}(t)\odot \bm{A}\right),\label{subequ:update}
\end{subnumcases}
where $t = 1,2,\ldots$, is the iteration index, $\bR(t)$ is an intermediate variable with $\bR(0) = \bzero$, $\mu$ is the step size, and $\mathcal{D}_\tau(\cdot)$ is the so-called shrinkage operator with the thresholding level $\tau$ \cite{caiSingularvalue2010}. Denote $\sigma_n(t), n = 1,\ldots,N,$ as the singular values of $\bR(t)$, and $\bU(t) \in\mathbb{R}^{N\times N}$ and $\bV(t) \in\mathbb{R}^{N\times N}$ as the left and right singular vector matrix, respectively. The shrinkage operator applied to $\bR(t)$ is defined as \cite{caiSingularvalue2010}
\begin{equation}\label{equ:shrinkage}
	\mathcal{D}_{\tau}\left(\bR(t)\right) = \sum_{n=1}^{N}[\sigma_n(t) - \tau]^+\bU(t)\bV(t)^\tT,
\end{equation}
where $[\sigma_n(t) - \tau]^+ = \max\left(0,\sigma_n(t)-\tau\right)$. The SVT algorithm terminates when the maximum number of iterations is reached, and returns $\widetilde{\bR}(t)$ as the estimation of the EDM $\bR$.

The update of $\bR(t)$ in \eqref{subequ:update} is performed elementwise and each node maintains one row of $\bR(t)$, which satisfies the assumption in Section \ref{sec:dSVDknonw}. Hence, the SVD of $\bR(t)$ can be realized via the d-raSVD1 algorithm distributedly. As a result, each node has access to the complete set of singular values, one row of $\bU(t)$, and the entire matrix $\bV(t)$. Thus, the shrinkage operation in \eqref{equ:shrinkage} is performed locally. Moreover, since only the product $\bU(t)\bV(t)^\tT$ rather than the matrix $\bV(t)$ itself is required to perform the shrinkage operation, the multiplication of $\bU(t)$ and $\bV(t)^\tT$ is carried out once a row of $\bV(t)$ is obtained via \eqref{equ:XUvn} to reduce the storage cost in each node. As \eqref{subequ:shrink} is carried out distributedly and \eqref{subequ:update} involves only local operation, \eqref{equ:svt} results in the d-SVT algorithm.

Noticing that the EDM $\bR(t)$ is real and symmetric, its left and right singular vectors associated with the same singular value are equal up to a real scaling factor, i.e., $\bu_n(t) = \delta_n(t)\bv_n(t)$ with $\delta_n(t) \in \{-1, 1\}, \forall n = 1,\ldots, N,$ denoting the scaling factor. In practice, the scaling factors can be determined by comparing the row of $\bU_n(t)$ that is available in the node with the corresponding row of $\bV_n(t)$. As a result, each node has complete knowledge of the SVD of $\bR(t)$. Hence, the classical MDS can be applied in each node locally. Specifically, the estimated coordinate matrix $\widetilde{\bX}$ is obtained via the principal subspace of the Gram matrix \cite{fanLocalizationsensor2024}
\begin{equation}\label{equ:Grank1}
	\bG = -\frac{1}{2}\sum_{n=1}^{N}\delta_{n}(t)[\sigma_{n}(t)-\tau]^+\widetilde{\bu}_{n}(t)\widetilde{\bu}_{n}(t)^\tT,
\end{equation}
where $\widetilde{\bu}_{n}(t) = \bu_{n}(t) - \bar{u}_{n}(t)\bone$ and $\bar{u}_{n}(t)$ is the average of all entries of $\bu_{n}(t)$. Mathematically, we have
\begin{equation}\label{equ:Xtildes}
	\begin{aligned}
		\widetilde{\bX}		&=[\psi_1^\frac{1}{2}\bm{f}_1,\ldots,\psi_h^\frac{1}{2}\bm{f}_h]^\tT,
	\end{aligned}
\end{equation}
where $\psi_n$ and $\bm{f}_n$, $n=1,\ldots,h,$ are the descendingly ordered eigenvalues and the associated eigenvectors, respectively, that are obtained by applying the rational function approximation approach to \eqref{equ:Grank1} in each node locally.

\begin{remark}
The prior information that $\bR(t)$ is real and symmetric can also be used to obtain the scaling factor $\delta_n(t)$ via consensus algorithms distributedly, and to obtain rows of $\bV_n(t)$ based on the corresponding rows of $\bU_n(t)$ \cite{fanLocalizationsensor2024}. This, however, leads to a decentralized MDS algorithm that requires more communication cost than the above proposed approach.
\end{remark}
\renewcommand{\arraystretch}{1.3}
\renewcommand*{\maxval}{0.35}
\begin{table*}
	\centering
	\caption{Simulation results of distributed localization for two dimensional sensor network.}
	\begin{tabular}{c|c||c|c||c|c||c|c}
		\multirow{2}{*}{\small{\makecell{Network size\\$N$}}} & \multirow{2}{*}{\small{\makecell{Percentage of\\missing entries ($\%$)}}} & \multicolumn{2}{c||}{d-pmSVD1} & \multicolumn{2}{c||}{d-raSVD1} & \multicolumn{2}{c}{d-TraEVD($\floor{\frac{N}{3}}$)} \\
		\cline{3-8} & & $\epsilon_{\bR}$ & $\epsilon_{\bX}$ & $\epsilon_{\bR}$ & $\epsilon_{\bX}$ & $\epsilon_{\bR}$ & $\epsilon_{\bX}$\\
		\hline\hline
		\multirow{3}{*}{$30$} & $10$ & $\gradient{0.0734}$ & $\gradient{0.0459}$ & $\gradient{0.0249}$ & $\gradient{0.0100}$ & $\gradient{0.0249}$ & $\gradient{0.0132}$\\
		\cline{2-8}
		& $20$& $\gradient{0.1387}$ & $\gradient{0.0799}$ & $\gradient{0.1051}$ & $\gradient{0.0521}$ & $\gradient{0.1091}$ & $\gradient{0.0792}$\\
		\cline{2-8}
		& $30$& $\gradient{0.2783}$ & $\gradient{0.1680}$ &  $\gradient{0.2637}$ & $\gradient{0.1506}$ & $\gradient{0.3128}$ & $\gradient{0.2379}$\\
		\hline
		\hline
		\multirow{3}{*}{$100$} & $10$ & $\gradient{0.0152}$ & $\gradient{0.0059}$ & $\gradient{0.0128}$ & $\gradient{0.0044}$ & $\gradient{0.0128}$ & $\gradient{0.0044}$\\
		\cline{2-8}
		& $20$& $\gradient{0.0513}$ & $\gradient{0.0257}$ & $\gradient{0.0507}$ & $\gradient{0.0253}$ & $\gradient{0.0507}$ & $\gradient{0.0269}$\\
		\cline{2-8}
		& $30$& $\gradient{0.1459}$ & $\gradient{0.0736}$ & $\gradient{0.1457}$ & $\gradient{0.0735}$ & $\gradient{0.1457}$ & $\gradient{0.0744}$\\
		\hline
	\end{tabular}
	\label{tab:disLocal}
\end{table*}
\renewcommand{\arraystretch}{1}
\paragraph{Decentralized reconstruction of the true coordinates} To reconstruct the true sensor coordinates, it is assumed that there are $N_\text{a}$ anchors. If the coordinates of $N_\text{a}$ anchors are available to all nodes, the true coordinates can be reconstructed in each node by following the approach in \cite{dokmanicEuclideandistance2015}. However, if the true coordinates are only available individually to the anchors, the following decentralized approach can be applied to estimate the true coordinates. 
Denoting the estimated coordinates of anchors as $\widetilde{\bX}_\text{a} \in\mathbb{R}^{h\times N_\text{a}}$, and the corresponding true coordinates of anchors as $\bX_\text{a}\in\mathbb{R}^{h\times N_\text{a}}$, the respective associated means $\tilde{\bx}_\text{a,c}$ and $\bx_\text{a,c}$ over all anchors are obtained via consensus algorithms distributedly. Hence, the coordinate transformation matrix $\bQ\in\mathbb{R}^{h\times h}$ is obtained as $\bQ = \bM\bK^\tT$, where
$\bK$ and $\bM$ are the left and right singular vector matrix of $\bC=\widetilde{\bX}_\text{a,c}\bX_\text{a,c}^\tT$, respectively, with $\widetilde{\bX}_\text{a,c} = \widetilde{\bX}_\text{a} - \tilde{\bx}_\text{a,c}\bone^\tT$ and $\bX_\text{a,c} = \bX_\text{a} - \bx_\text{a,c}\bone^\tT$ \cite{dokmanicEuclideandistance2015,schoenemannSolutionorthogonal1964}.
\begin{figure}[t]
	\begin{subfigure}{.47\columnwidth}
		\centering
		\include{simulations/network_localization}
		\caption{}
		\label{subfig:localNet}
	\end{subfigure}
	\hfill
	\begin{subfigure}{.47\columnwidth}
		\centering
		\include{simulations/result_localization}
		\caption{}
		\label{subfig:localResult}
	\end{subfigure}
	\caption{(a): A sensor network with $N=30$ nodes (black dots) and $6$ obstacles (blue circles). The percentage of missing entries in the EDM is $22.5\%$. (b): Estimation of the true coordinates (black dots) using the d-raSVD1 (red $+$) and the d-pmSVD1 (blue $\star$). Recovery errors are $\epsilon_{\bR,\text{d-EVD}} = 0.0858$, $\epsilon_{\bR,\text{d-PM}} = 0.1060$, $\epsilon_{\bX,\text{d-EVD}} = 0.0447$ and  $\epsilon_{\bX,\text{d-PM}} = 0.0539$.}
\end{figure}

To make the SVD of $\bC$ available to all nodes, we construct two sparse matrices $\bX_\text{c}\in\mathbb{R}^{h\times N}$ and $\widetilde{\bX}_\text{c}\in\mathbb{R}^{h\times N}$ whose $i$th columns are the corresponding columns of $\bX_\text{a,c}$ and of $\widetilde{\bX}_\text{a,c}$, respectively, that are associated with the $i$th node. Hence, entries of $\bC$ are obtained distributedly by
\begin{equation}\label{equ:rij}
	c_{ij} = \bar{\tilde{\bx}}_{\text{c},i}^\tT\bar{\bx}_{\text{c},j} = \ps{\bar{\tilde{\bx}}_{\text{c},i}\odot\bar{\bx}_{\text{c},j}}, \ i,j=1,\ldots,h,
\end{equation}
where $\bar{\tilde{\bx}}_{\text{c},i}$ and $\bar{\bx}_{\text{c},j}$ are the $i$th row of $\widetilde{\bX}_\text{c}$ and the $j$th row of $\bX_\text{c}$, respectively. 
Since the dimension of $\bC$ is $h\times h$ and $h$ is usually small, the communication cost to compute all entries of $\bC$ as well as the computational cost associated with the SVD of $\bC$ in each node is negligible. After $\bQ$ is available to all sensors, the estimated true coordinates ${\bX}_\text{est}= [{\bx}_{1,\text{est}},\ldots,{\bx}_{N,\text{est}}]$ are reconstructed locally by
\begin{equation}
	\bx_{i,\text{est}} = \bQ(\tilde{\bx} - \tilde{\bx}_\text{a,c}) + \bx_\text{a,c},\quad i = 1,\ldots,N.
\end{equation}
\paragraph{Experimental results} Random sensor networks and obstacles that block a portion of LoS communication links in the two dimensional vector space, i.e., $h=2$, are generated, where only the incomplete EDM is available\footnote{In this application, we assume that the inter-sensor distances are ideally estimated (i.e., we consider the noise-free case) and focus on reconstruction from an incomplete EDM obtained in the presence of obstacles that block LoS transmissions among nodes resulting in missing distance measurements for certain links. In a more practical scenario, the performance of the proposed algorithms is also affected by measurement noise.}. One example of such networks is depicted in Fig.~\ref{subfig:localNet}. The proposed decentralized localization approach is then applied to estimate the true coordinates of sensors, where the maximum iteration number and the step size of the d-SVT algorithm are $200$ and $\mu = 1.5$, respectively. Moreover, the thresholding level of the shrinkage operator $\tau=5N$, the maximum number of PS iterations is $N$, and $N_\text{a} = 5$ anchors are applied. As a comparison, the d-pmSVD1 approach with PM iteration number $P=4N$, and the d-TraEVD($d$) approach with $d=\floor{N/3}$ and $d=\floor{N/2}$ are applied, where $\floor{\cdot}$ rounds a number down to the nearest integer. The estimation results corresponding to the network in Fig.~\ref{subfig:localNet} is illustrated in Fig.~\ref{subfig:localResult}.

The recovery error between the estimated and the true EDMs and between the estimated and the true coordinates are defined as $	\epsilon_{\bR} = {\|{\widetilde{\bR}(t)} - \bR\|_\text{F}}/{\|\bR\|_\text{F}}$ and $\epsilon_{\bX} = {\left\|{\bX_\text{est}} - \bX\right\|_\text{F}}/{\left\|\bX\right\|_\text{F}}$, 
respectively. The simulation results averaged over $100$ random network realizations are summarized in Table~\ref{tab:disLocal}. It is observed that the proposed d-raSVD1 approach has a better error performance compared to the d-pmSVD1 approach for both $N=30$ and $N=100$ under different sparsity setups of the EDM. In addition, while the d-raSVD1 approach requires $N^2$ consensus instances in each d-SVT algorithm, the d-pmSVD1 approach needs already $4N^2+N+2$ consensus instances to compute only the largest singular value. 
From Table~\ref{tab:disLocal}, it is also observed that the d-TraEVD($d$) algorithm achieves similar error performance as the d-raSVD1 algorithm with the truncation dimension $d = \floor{\frac{N}{3}}$, which indicates that the associated communication cost can be further reduced by $66\%$ for the sensor localization application.

%% file: simulations/network_localization.tex
\begin{tikzpicture}[scale=1.3]
\def\obstacle#1#2{
	\node[circle,very thin,draw=blue,fill=yellow!30!white,inner sep=0,minimum size=1.8mm](#1) at #2 {};}
	\foreach \i/\x in {
		{o1/(1.4389cm,1.5667cm)},
		{o2/(2.0728cm,2.2805cm)},
		{o3/(1.3938cm,1.8193cm)},
		{o4/(1.0024cm,2.1341cm)},
		{o5/(2.5337cm,2.4756cm)},
		{o6/(2.3906cm,1.2648cm)}}
	\obstacle{\i}{\x};
\def\sensor#1#2{
	\node[circle,draw=black,fill=black,inner sep=0,minimum size=2pt](#1) at #2 {};}
	\foreach \i/\x in {
		{s1/(0.5765cm,2.0753cm)},
		{s2/(2.8599cm,1.5245cm)},
		{s3/(0.7669cm,1.9823cm)},
		{s4/(1.2475cm,1.0957cm)},
		{s5/(1.6201cm,0.5421cm)},
		{s6/(0.6936cm,0.3690cm)},
		{s7/(0.4143cm,2.1474cm)},
		{s8/(0.4713cm,1.9268cm)},
		{s9/(0.6198cm,2.9896cm)},
		{s10/(0.6938cm,2.0918cm)},
		{s11/(0.4164cm,0.5795cm)},
		{s12/(0.1854cm,0.1939cm)},
		{s13/(1.9774cm,2.1185cm)},
		{s14/(1.2331cm,0.1859cm)},
		{s15/(1.9031cm,2.4523cm)},
		{s16/(0.8084cm,2.8066cm)},
		{s17/(2.9797cm,1.6147cm)},
		{s18/(0.1317cm,0.7420cm)},
		{s19/(0.8302cm,0.2716cm)},
		{s20/(2.9561cm,2.7646cm)},
		{s21/(1.6918cm,2.2334cm)},
		{s22/(1.0725cm,0.4454cm)},
		{s23/(2.2809cm,1.6492cm)},
		{s24/(1.3106cm,0.6801cm)},
		{s25/(1.8567cm,2.2192cm)},
		{s26/(2.0364cm,1.5237cm)},
		{s27/(0.2097cm,2.5785cm)},
		{s28/(0.1919cm,1.8774cm)},
		{s29/(1.4385cm,2.2864cm)},
		{s30/(1.9803cm,0.6506cm)},
		{s30/(1.9803cm,0.6506cm)}}
	\sensor{\i}{\x};
	\begin{scope}[on background layer]
		\draw[ultra thin,red] (s1) -- (s3);
		\draw[ultra thin,red] (s1) -- (s4);
		\draw[ultra thin,red] (s1) -- (s5);
		\draw[ultra thin,red] (s1) -- (s6);
		\draw[ultra thin,red] (s1) -- (s7);
		\draw[ultra thin,red] (s1) -- (s8);
		\draw[ultra thin,red] (s1) -- (s9);
		\draw[ultra thin,red] (s1) -- (s10);
		\draw[ultra thin,red] (s1) -- (s11);
		\draw[ultra thin,red] (s1) -- (s12);
		\draw[ultra thin,red] (s1) -- (s14);
		\draw[ultra thin,red] (s1) -- (s16);
		\draw[ultra thin,red] (s1) -- (s17);
		\draw[ultra thin,red] (s1) -- (s18);
		\draw[ultra thin,red] (s1) -- (s19);
		\draw[ultra thin,red] (s1) -- (s22);
		\draw[ultra thin,red] (s1) -- (s24);
		\draw[ultra thin,red] (s1) -- (s27);
		\draw[ultra thin,red] (s1) -- (s28);
		\draw[ultra thin,red] (s1) -- (s30);
		\draw[ultra thin,red] (s2) -- (s4);
		\draw[ultra thin,red] (s2) -- (s9);
		\draw[ultra thin,red] (s2) -- (s13);
		\draw[ultra thin,red] (s2) -- (s14);
		\draw[ultra thin,red] (s2) -- (s16);
		\draw[ultra thin,red] (s2) -- (s17);
		\draw[ultra thin,red] (s2) -- (s18);
		\draw[ultra thin,red] (s2) -- (s20);
		\draw[ultra thin,red] (s2) -- (s21);
		\draw[ultra thin,red] (s2) -- (s23);
		\draw[ultra thin,red] (s2) -- (s25);
		\draw[ultra thin,red] (s2) -- (s26);
		\draw[ultra thin,red] (s2) -- (s27);
		\draw[ultra thin,red] (s2) -- (s28);
		\draw[ultra thin,red] (s2) -- (s29);
		\draw[ultra thin,red] (s2) -- (s30);
		\draw[ultra thin,red] (s3) -- (s4);
		\draw[ultra thin,red] (s3) -- (s5);
		\draw[ultra thin,red] (s3) -- (s6);
		\draw[ultra thin,red] (s3) -- (s7);
		\draw[ultra thin,red] (s3) -- (s8);
		\draw[ultra thin,red] (s3) -- (s9);
		\draw[ultra thin,red] (s3) -- (s10);
		\draw[ultra thin,red] (s3) -- (s11);
		\draw[ultra thin,red] (s3) -- (s12);
		\draw[ultra thin,red] (s3) -- (s13);
		\draw[ultra thin,red] (s3) -- (s14);
		\draw[ultra thin,red] (s3) -- (s16);
		\draw[ultra thin,red] (s3) -- (s18);
		\draw[ultra thin,red] (s3) -- (s19);
		\draw[ultra thin,red] (s3) -- (s22);
		\draw[ultra thin,red] (s3) -- (s24);
		\draw[ultra thin,red] (s3) -- (s25);
		\draw[ultra thin,red] (s3) -- (s27);
		\draw[ultra thin,red] (s3) -- (s28);
		\draw[ultra thin,red] (s3) -- (s30);
		\draw[ultra thin,red] (s4) -- (s5);
		\draw[ultra thin,red] (s4) -- (s6);
		\draw[ultra thin,red] (s4) -- (s7);
		\draw[ultra thin,red] (s4) -- (s8);
		\draw[ultra thin,red] (s4) -- (s9);
		\draw[ultra thin,red] (s4) -- (s10);
		\draw[ultra thin,red] (s4) -- (s11);
		\draw[ultra thin,red] (s4) -- (s12);
		\draw[ultra thin,red] (s4) -- (s13);
		\draw[ultra thin,red] (s4) -- (s14);
		\draw[ultra thin,red] (s4) -- (s17);
		\draw[ultra thin,red] (s4) -- (s18);
		\draw[ultra thin,red] (s4) -- (s19);
		\draw[ultra thin,red] (s4) -- (s22);
		\draw[ultra thin,red] (s4) -- (s23);
		\draw[ultra thin,red] (s4) -- (s24);
		\draw[ultra thin,red] (s4) -- (s26);
		\draw[ultra thin,red] (s4) -- (s27);
		\draw[ultra thin,red] (s4) -- (s28);
		\draw[ultra thin,red] (s4) -- (s30);
		\draw[ultra thin,red] (s5) -- (s6);
		\draw[ultra thin,red] (s5) -- (s7);
		\draw[ultra thin,red] (s5) -- (s8);
		\draw[ultra thin,red] (s5) -- (s10);
		\draw[ultra thin,red] (s5) -- (s11);
		\draw[ultra thin,red] (s5) -- (s12);
		\draw[ultra thin,red] (s5) -- (s13);
		\draw[ultra thin,red] (s5) -- (s14);
		\draw[ultra thin,red] (s5) -- (s15);
		\draw[ultra thin,red] (s5) -- (s17);
		\draw[ultra thin,red] (s5) -- (s18);
		\draw[ultra thin,red] (s5) -- (s19);
		\draw[ultra thin,red] (s5) -- (s20);
		\draw[ultra thin,red] (s5) -- (s21);
		\draw[ultra thin,red] (s5) -- (s22);
		\draw[ultra thin,red] (s5) -- (s23);
		\draw[ultra thin,red] (s5) -- (s24);
		\draw[ultra thin,red] (s5) -- (s25);
		\draw[ultra thin,red] (s5) -- (s26);
		\draw[ultra thin,red] (s5) -- (s27);
		\draw[ultra thin,red] (s5) -- (s28);
		\draw[ultra thin,red] (s5) -- (s30);
		\draw[ultra thin,red] (s6) -- (s7);
		\draw[ultra thin,red] (s6) -- (s8);
		\draw[ultra thin,red] (s6) -- (s9);
		\draw[ultra thin,red] (s6) -- (s10);
		\draw[ultra thin,red] (s6) -- (s11);
		\draw[ultra thin,red] (s6) -- (s12);
		\draw[ultra thin,red] (s6) -- (s13);
		\draw[ultra thin,red] (s6) -- (s14);
		\draw[ultra thin,red] (s6) -- (s16);
		\draw[ultra thin,red] (s6) -- (s18);
		\draw[ultra thin,red] (s6) -- (s19);
		\draw[ultra thin,red] (s6) -- (s20);
		\draw[ultra thin,red] (s6) -- (s22);
		\draw[ultra thin,red] (s6) -- (s23);
		\draw[ultra thin,red] (s6) -- (s24);
		\draw[ultra thin,red] (s6) -- (s26);
		\draw[ultra thin,red] (s6) -- (s27);
		\draw[ultra thin,red] (s6) -- (s28);
		\draw[ultra thin,red] (s6) -- (s29);
		\draw[ultra thin,red] (s6) -- (s30);
		\draw[ultra thin,red] (s7) -- (s8);
		\draw[ultra thin,red] (s7) -- (s9);
		\draw[ultra thin,red] (s7) -- (s10);
		\draw[ultra thin,red] (s7) -- (s11);
		\draw[ultra thin,red] (s7) -- (s12);
		\draw[ultra thin,red] (s7) -- (s14);
		\draw[ultra thin,red] (s7) -- (s15);
		\draw[ultra thin,red] (s7) -- (s16);
		\draw[ultra thin,red] (s7) -- (s17);
		\draw[ultra thin,red] (s7) -- (s18);
		\draw[ultra thin,red] (s7) -- (s19);
		\draw[ultra thin,red] (s7) -- (s20);
		\draw[ultra thin,red] (s7) -- (s22);
		\draw[ultra thin,red] (s7) -- (s24);
		\draw[ultra thin,red] (s7) -- (s27);
		\draw[ultra thin,red] (s7) -- (s28);
		\draw[ultra thin,red] (s7) -- (s29);
		\draw[ultra thin,red] (s7) -- (s30);
		\draw[ultra thin,red] (s8) -- (s9);
		\draw[ultra thin,red] (s8) -- (s10);
		\draw[ultra thin,red] (s8) -- (s11);
		\draw[ultra thin,red] (s8) -- (s12);
		\draw[ultra thin,red] (s8) -- (s13);
		\draw[ultra thin,red] (s8) -- (s14);
		\draw[ultra thin,red] (s8) -- (s16);
		\draw[ultra thin,red] (s8) -- (s18);
		\draw[ultra thin,red] (s8) -- (s19);
		\draw[ultra thin,red] (s8) -- (s22);
		\draw[ultra thin,red] (s8) -- (s24);
		\draw[ultra thin,red] (s8) -- (s25);
		\draw[ultra thin,red] (s8) -- (s26);
		\draw[ultra thin,red] (s8) -- (s27);
		\draw[ultra thin,red] (s8) -- (s28);
		\draw[ultra thin,red] (s8) -- (s30);
		\draw[ultra thin,red] (s9) -- (s10);
		\draw[ultra thin,red] (s9) -- (s11);
		\draw[ultra thin,red] (s9) -- (s12);
		\draw[ultra thin,red] (s9) -- (s13);
		\draw[ultra thin,red] (s9) -- (s14);
		\draw[ultra thin,red] (s9) -- (s15);
		\draw[ultra thin,red] (s9) -- (s16);
		\draw[ultra thin,red] (s9) -- (s17);
		\draw[ultra thin,red] (s9) -- (s18);
		\draw[ultra thin,red] (s9) -- (s19);
		\draw[ultra thin,red] (s9) -- (s20);
		\draw[ultra thin,red] (s9) -- (s21);
		\draw[ultra thin,red] (s9) -- (s22);
		\draw[ultra thin,red] (s9) -- (s23);
		\draw[ultra thin,red] (s9) -- (s24);
		\draw[ultra thin,red] (s9) -- (s25);
		\draw[ultra thin,red] (s9) -- (s26);
		\draw[ultra thin,red] (s9) -- (s27);
		\draw[ultra thin,red] (s9) -- (s28);
		\draw[ultra thin,red] (s9) -- (s29);
		\draw[ultra thin,red] (s10) -- (s11);
		\draw[ultra thin,red] (s10) -- (s12);
		\draw[ultra thin,red] (s10) -- (s14);
		\draw[ultra thin,red] (s10) -- (s16);
		\draw[ultra thin,red] (s10) -- (s17);
		\draw[ultra thin,red] (s10) -- (s18);
		\draw[ultra thin,red] (s10) -- (s19);
		\draw[ultra thin,red] (s10) -- (s22);
		\draw[ultra thin,red] (s10) -- (s24);
		\draw[ultra thin,red] (s10) -- (s27);
		\draw[ultra thin,red] (s10) -- (s28);
		\draw[ultra thin,red] (s10) -- (s30);
		\draw[ultra thin,red] (s11) -- (s12);
		\draw[ultra thin,red] (s11) -- (s14);
		\draw[ultra thin,red] (s11) -- (s16);
		\draw[ultra thin,red] (s11) -- (s17);
		\draw[ultra thin,red] (s11) -- (s18);
		\draw[ultra thin,red] (s11) -- (s19);
		\draw[ultra thin,red] (s11) -- (s22);
		\draw[ultra thin,red] (s11) -- (s23);
		\draw[ultra thin,red] (s11) -- (s24);
		\draw[ultra thin,red] (s11) -- (s26);
		\draw[ultra thin,red] (s11) -- (s27);
		\draw[ultra thin,red] (s11) -- (s28);
		\draw[ultra thin,red] (s11) -- (s29);
		\draw[ultra thin,red] (s11) -- (s30);
		\draw[ultra thin,red] (s12) -- (s14);
		\draw[ultra thin,red] (s12) -- (s16);
		\draw[ultra thin,red] (s12) -- (s18);
		\draw[ultra thin,red] (s12) -- (s19);
		\draw[ultra thin,red] (s12) -- (s22);
		\draw[ultra thin,red] (s12) -- (s23);
		\draw[ultra thin,red] (s12) -- (s24);
		\draw[ultra thin,red] (s12) -- (s25);
		\draw[ultra thin,red] (s12) -- (s26);
		\draw[ultra thin,red] (s12) -- (s27);
		\draw[ultra thin,red] (s12) -- (s28);
		\draw[ultra thin,red] (s12) -- (s29);
		\draw[ultra thin,red] (s12) -- (s30);
		\draw[ultra thin,red] (s13) -- (s14);
		\draw[ultra thin,red] (s13) -- (s15);
		\draw[ultra thin,red] (s13) -- (s16);
		\draw[ultra thin,red] (s13) -- (s17);
		\draw[ultra thin,red] (s13) -- (s18);
		\draw[ultra thin,red] (s13) -- (s19);
		\draw[ultra thin,red] (s13) -- (s21);
		\draw[ultra thin,red] (s13) -- (s22);
		\draw[ultra thin,red] (s13) -- (s23);
		\draw[ultra thin,red] (s13) -- (s24);
		\draw[ultra thin,red] (s13) -- (s25);
		\draw[ultra thin,red] (s13) -- (s26);
		\draw[ultra thin,red] (s13) -- (s27);
		\draw[ultra thin,red] (s13) -- (s28);
		\draw[ultra thin,red] (s13) -- (s29);
		\draw[ultra thin,red] (s13) -- (s30);
		\draw[ultra thin,red] (s14) -- (s15);
		\draw[ultra thin,red] (s14) -- (s17);
		\draw[ultra thin,red] (s14) -- (s18);
		\draw[ultra thin,red] (s14) -- (s19);
		\draw[ultra thin,red] (s14) -- (s20);
		\draw[ultra thin,red] (s14) -- (s21);
		\draw[ultra thin,red] (s14) -- (s22);
		\draw[ultra thin,red] (s14) -- (s23);
		\draw[ultra thin,red] (s14) -- (s24);
		\draw[ultra thin,red] (s14) -- (s25);
		\draw[ultra thin,red] (s14) -- (s26);
		\draw[ultra thin,red] (s14) -- (s27);
		\draw[ultra thin,red] (s14) -- (s28);
		\draw[ultra thin,red] (s14) -- (s30);
		\draw[ultra thin,red] (s15) -- (s16);
		\draw[ultra thin,red] (s15) -- (s18);
		\draw[ultra thin,red] (s15) -- (s20);
		\draw[ultra thin,red] (s15) -- (s21);
		\draw[ultra thin,red] (s15) -- (s22);
		\draw[ultra thin,red] (s15) -- (s24);
		\draw[ultra thin,red] (s15) -- (s25);
		\draw[ultra thin,red] (s15) -- (s26);
		\draw[ultra thin,red] (s15) -- (s27);
		\draw[ultra thin,red] (s15) -- (s29);
		\draw[ultra thin,red] (s15) -- (s30);
		\draw[ultra thin,red] (s16) -- (s17);
		\draw[ultra thin,red] (s16) -- (s18);
		\draw[ultra thin,red] (s16) -- (s19);
		\draw[ultra thin,red] (s16) -- (s20);
		\draw[ultra thin,red] (s16) -- (s21);
		\draw[ultra thin,red] (s16) -- (s22);
		\draw[ultra thin,red] (s16) -- (s23);
		\draw[ultra thin,red] (s16) -- (s25);
		\draw[ultra thin,red] (s16) -- (s26);
		\draw[ultra thin,red] (s16) -- (s27);
		\draw[ultra thin,red] (s16) -- (s28);
		\draw[ultra thin,red] (s16) -- (s29);
		\draw[ultra thin,red] (s17) -- (s18);
		\draw[ultra thin,red] (s17) -- (s20);
		\draw[ultra thin,red] (s17) -- (s21);
		\draw[ultra thin,red] (s17) -- (s23);
		\draw[ultra thin,red] (s17) -- (s25);
		\draw[ultra thin,red] (s17) -- (s26);
		\draw[ultra thin,red] (s17) -- (s27);
		\draw[ultra thin,red] (s17) -- (s29);
		\draw[ultra thin,red] (s17) -- (s30);
		\draw[ultra thin,red] (s18) -- (s19);
		\draw[ultra thin,red] (s18) -- (s21);
		\draw[ultra thin,red] (s18) -- (s22);
		\draw[ultra thin,red] (s18) -- (s23);
		\draw[ultra thin,red] (s18) -- (s24);
		\draw[ultra thin,red] (s18) -- (s26);
		\draw[ultra thin,red] (s18) -- (s27);
		\draw[ultra thin,red] (s18) -- (s28);
		\draw[ultra thin,red] (s18) -- (s29);
		\draw[ultra thin,red] (s18) -- (s30);
		\draw[ultra thin,red] (s19) -- (s20);
		\draw[ultra thin,red] (s19) -- (s22);
		\draw[ultra thin,red] (s19) -- (s23);
		\draw[ultra thin,red] (s19) -- (s24);
		\draw[ultra thin,red] (s19) -- (s26);
		\draw[ultra thin,red] (s19) -- (s27);
		\draw[ultra thin,red] (s19) -- (s28);
		\draw[ultra thin,red] (s19) -- (s29);
		\draw[ultra thin,red] (s19) -- (s30);
		\draw[ultra thin,red] (s20) -- (s21);
		\draw[ultra thin,red] (s20) -- (s22);
		\draw[ultra thin,red] (s20) -- (s23);
		\draw[ultra thin,red] (s20) -- (s24);
		\draw[ultra thin,red] (s20) -- (s26);
		\draw[ultra thin,red] (s20) -- (s27);
		\draw[ultra thin,red] (s20) -- (s29);
		\draw[ultra thin,red] (s20) -- (s30);
		\draw[ultra thin,red] (s21) -- (s23);
		\draw[ultra thin,red] (s21) -- (s25);
		\draw[ultra thin,red] (s21) -- (s26);
		\draw[ultra thin,red] (s21) -- (s27);
		\draw[ultra thin,red] (s21) -- (s29);
		\draw[ultra thin,red] (s21) -- (s30);
		\draw[ultra thin,red] (s22) -- (s23);
		\draw[ultra thin,red] (s22) -- (s24);
		\draw[ultra thin,red] (s22) -- (s25);
		\draw[ultra thin,red] (s22) -- (s26);
		\draw[ultra thin,red] (s22) -- (s27);
		\draw[ultra thin,red] (s22) -- (s28);
		\draw[ultra thin,red] (s22) -- (s30);
		\draw[ultra thin,red] (s23) -- (s24);
		\draw[ultra thin,red] (s23) -- (s25);
		\draw[ultra thin,red] (s23) -- (s26);
		\draw[ultra thin,red] (s23) -- (s29);
		\draw[ultra thin,red] (s23) -- (s30);
		\draw[ultra thin,red] (s24) -- (s25);
		\draw[ultra thin,red] (s24) -- (s26);
		\draw[ultra thin,red] (s24) -- (s27);
		\draw[ultra thin,red] (s24) -- (s28);
		\draw[ultra thin,red] (s24) -- (s30);
		\draw[ultra thin,red] (s25) -- (s26);
		\draw[ultra thin,red] (s25) -- (s27);
		\draw[ultra thin,red] (s25) -- (s29);
		\draw[ultra thin,red] (s25) -- (s30);
		\draw[ultra thin,red] (s26) -- (s29);
		\draw[ultra thin,red] (s26) -- (s30);
		\draw[ultra thin,red] (s27) -- (s28);
		\draw[ultra thin,red] (s27) -- (s29);
		\draw[ultra thin,red] (s27) -- (s30);
		\draw[ultra thin,red] (s28) -- (s30);
		\draw[ultra thin,red] (s29) -- (s30);
	\end{scope}
\end{tikzpicture}

%% file: simulations/result_localization.tex
\begin{tikzpicture}[scale=1.3]
\def\sensor#1#2{
	\node[circle,draw=black,fill=black,inner sep=0,minimum size=2pt](#1) at #2 {};}
	\foreach \i/\x in {
		{s1/(0.5765cm,2.0753cm)},
		{s2/(2.8599cm,1.5245cm)},
		{s3/(0.7669cm,1.9823cm)},
		{s4/(1.2475cm,1.0957cm)},
		{s5/(1.6201cm,0.5421cm)},
		{s6/(0.6936cm,0.3690cm)},
		{s7/(0.4143cm,2.1474cm)},
		{s8/(0.4713cm,1.9268cm)},
		{s9/(0.6198cm,2.9896cm)},
		{s10/(0.6938cm,2.0918cm)},
		{s11/(0.4164cm,0.5795cm)},
		{s12/(0.1854cm,0.1939cm)},
		{s13/(1.9774cm,2.1185cm)},
		{s14/(1.2331cm,0.1859cm)},
		{s15/(1.9031cm,2.4523cm)},
		{s16/(0.8084cm,2.8066cm)},
		{s17/(2.9797cm,1.6147cm)},
		{s18/(0.1317cm,0.7420cm)},
		{s19/(0.8302cm,0.2716cm)},
		{s20/(2.9561cm,2.7646cm)},
		{s21/(1.6918cm,2.2334cm)},
		{s22/(1.0725cm,0.4454cm)},
		{s23/(2.2809cm,1.6492cm)},
		{s24/(1.3106cm,0.6801cm)},
		{s25/(1.8567cm,2.2192cm)},
		{s26/(2.0364cm,1.5237cm)},
		{s27/(0.2097cm,2.5785cm)},
		{s28/(0.1919cm,1.8774cm)},
		{s29/(1.4385cm,2.2864cm)},
		{s30/(1.9803cm,0.6506cm)},
		{s30/(1.9803cm,0.6506cm)}}
	\sensor{\i}{\x};
\def\result#1#2{
	\node(#1) at #2 {\textcolor{red}{\footnotesize{+}}};}
	\foreach \i/\x in {
		{s1/(0.5422cm,2.0526cm)},
		{s2/(2.8077cm,1.5425cm)},
		{s3/(0.8685cm,1.9934cm)},
		{s4/(1.2365cm,1.0964cm)},
		{s5/(1.6161cm,0.5351cm)},
		{s6/(0.6605cm,0.3417cm)},
		{s7/(0.3483cm,2.1101cm)},
		{s8/(0.5846cm,1.9342cm)},
		{s9/(0.5588cm,2.9576cm)},
		{s10/(0.6600cm,2.0711cm)},
		{s11/(0.4162cm,0.5587cm)},
		{s12/(0.3679cm,0.2094cm)},
		{s13/(1.9654cm,2.1101cm)},
		{s14/(1.2277cm,0.1661cm)},
		{s15/(1.8379cm,2.4395cm)},
		{s16/(0.7551cm,2.7791cm)},
		{s17/(2.9179cm,1.6437cm)},
		{s18/(0.1035cm,0.7019cm)},
		{s19/(0.8042cm,0.2478cm)},
		{s20/(2.8076cm,2.7844cm)},
		{s21/(1.6492cm,2.2157cm)},
		{s22/(1.0472cm,0.4284cm)},
		{s23/(2.2725cm,1.6637cm)},
		{s24/(1.2829cm,0.6719cm)},
		{s25/(1.8664cm,2.2341cm)},
		{s26/(2.0349cm,1.5263cm)},
		{s27/(0.1429cm,2.5349cm)},
		{s28/(0.1573cm,1.8441cm)},
		{s29/(1.4087cm,2.2759cm)},
		{s30/(1.9843cm,0.6505cm)},
		{s30/(1.9843cm,0.6505cm)}}
	\result{\i}{\x};
\def\result#1#2{
	\node(#1) at #2 {\textcolor{blue}{\footnotesize{$\star$}}};}
	\foreach \i/\x in {
		{s1/(0.4683cm,2.0648cm)},
		{s2/(2.8196cm,1.4776cm)},
		{s3/(0.8600cm,2.0097cm)},
		{s4/(1.2432cm,1.1046cm)},
		{s5/(1.6799cm,0.5632cm)},
		{s6/(0.6748cm,0.3989cm)},
		{s7/(0.2254cm,2.1182cm)},
		{s8/(0.5452cm,1.9646cm)},
		{s9/(0.3913cm,2.9360cm)},
		{s10/(0.5880cm,2.0772cm)},
		{s11/(0.4198cm,0.6275cm)},
		{s12/(0.3664cm,0.3130cm)},
		{s13/(1.9592cm,2.0695cm)},
		{s14/(1.2740cm,0.2083cm)},
		{s15/(1.7450cm,2.3813cm)},
		{s16/(0.6323cm,2.7489cm)},
		{s17/(2.9383cm,1.6069cm)},
		{s18/(0.0160cm,0.7535cm)},
		{s19/(0.8346cm,0.3042cm)},
		{s20/(2.6553cm,2.7164cm)},
		{s21/(1.5872cm,2.1646cm)},
		{s22/(1.0841cm,0.4663cm)},
		{s23/(2.3071cm,1.6375cm)},
		{s24/(1.3319cm,0.6980cm)},
		{s25/(1.8700cm,2.2047cm)},
		{s26/(2.0836cm,1.5061cm)},
		{s27/(-0.0213cm,2.5380cm)},
		{s28/(0.0613cm,1.8797cm)},
		{s29/(1.3730cm,2.2529cm)},
		{s30/(2.0657cm,0.6627cm)},
		{s30/(2.0657cm,0.6627cm)}}
	\result{\i}{\x};
\end{tikzpicture}

%% file: simulations/passiveRadar1.tex
	\begin{tikzpicture}[scale=1.5]
		
		\filldraw [fill=black!10] (-7.,5) -- (-7.,-0.1) -- (1.5,-0.1)[rounded corners] -- (1.5,1.)[rounded corners] -- (-1,2.3)[rounded corners] -- (-3,3.5)[rounded corners] -- (-5,4.8)[rounded corners] -- (-7,5);

		\node[inner sep=0pt] (bs1) at (-6.,5) {\includegraphics[width=.035\textwidth]{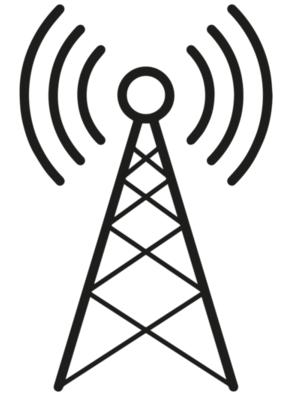}};
		\node (bs) at (-6.1,4.) [text width=2cm, align=center]{\scriptsize{Non-cooperative\\ [-1ex]Illuminator}};
				
		\node[inner sep=0pt] (aircraft) at (-1.5,5.5) {\huge{\Plane}};
		\node at (-.7,5.5) {\scriptsize{Target}};
		
		\draw [decorate,decoration={expanding waves,angle=5}](bs1) - -	(aircraft) ;
		
		\newcommand\myline[1][]{%
			\,\tikz[baseline]\draw[very thick,#1](0,-\dp\strutbox)--(0,\ht\strutbox);\,%
		}
		\newcommand{\myNode}{
				\scalebox{-1}[1]{\includegraphics[width=.04\textwidth]{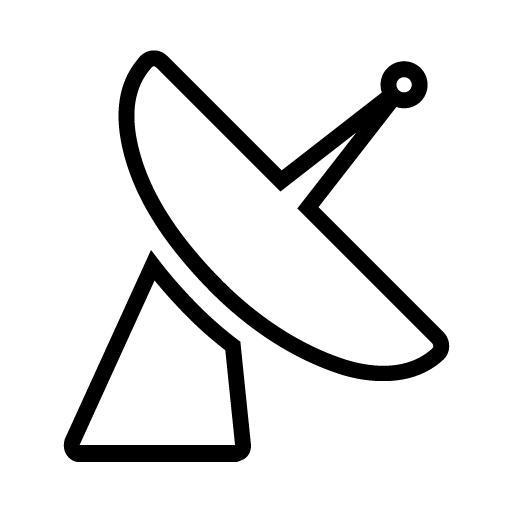}}\hspace*{-1mm}\includegraphics[width=.04\textwidth]{./simulations/radar.png}
		}
		
		\node[inner sep=0pt] (node1) at (-6.2,1.6) {\myNode};
		\node[inner sep=0pt] (node2) at (-3.7,2.5) {\myNode};
		\node[inner sep=0pt] (node3) at (-4.2,0.5) {\myNode};
		\node[inner sep=0pt] (node4) at (-1.8,0.3) {\myNode};
		\node[inner sep=0pt] (node5) at (-1.1,2.1) {\myNode};
		\node[inner sep=0pt] (node6) at (.7,.78) {\myNode};
		\node[rotate=-49] (etc) at (-2.7,1.4) {$\boldsymbol{\cdots}$};
		
		\draw[thick] (node1) -- (node2);
		\draw[thick] (node1) -- (node3);
		\draw[thick] (node2) -- (node3);
		\draw[thick] (node3) -- (node4);
		
		\draw[thick] (node2) -- (etc);
		\draw[thick] (node4) -- (etc);
		
		\draw[thick] (node2) -- (node5);
		\draw[thick] (node4) -- (node5);
		\draw[thick] (node4) -- (node6);
		\draw[thick] (node5) -- (node6);
		
		\node[inner sep=0pt] (refCh) at (-6.4,0.4) [text width=1.cm, align=center]{\scriptsize{Reference\\[-1ex]channel}};
		\draw[-latex,very thin,dotted] (refCh) -- (refCh.north|-node1.south);
		\draw[-latex,very thin,dotted] (refCh.east) -- (node3.west|-refCh.east);
		
		\node[inner sep=0pt] (surCh) at (1.03,2.2) [text width=1.23cm, align=center]{\scriptsize{Surveillance\\[-1ex]channel}};
		\draw[-latex,very thin,dotted] (surCh) -- (surCh.south|-node6.north);
		\draw[-latex,very thin,dotted] (surCh) -- (node5.east|-surCh.west);
		\draw [decorate,decoration={expanding waves,angle=15}](bs1) -- (node2) ;
		
		\node at (-5.5,3.1) {\scriptsize{Direct path}};
		
		\draw [decorate,decoration={expanding waves,angle=3}](aircraft) -- (node2.35) ;
		\node at (-1.2,4.2) {\scriptsize{Reflection path}};

		

		

	\end{tikzpicture}

%% file: simulations/passiveRadarSimulation_new.tex
%
%
\definecolor{mycolor1}{rgb}{0.00000,0.44700,0.74100}%
\definecolor{mycolor2}{rgb}{0.85000,0.32500,0.09800}%
\definecolor{mycolor3}{rgb}{0.92900,0.69400,0.12500}%
\definecolor{mycolor4}{rgb}{0.49400,0.18400,0.55600}%
\definecolor{mycolor5}{rgb}{0.46600,0.67400,0.18800}%
\begin{tikzpicture}

\begin{axis}[%
width=7cm,
height=4cm,
scale only axis,
xmin=0,
xmax=1,
ymin=0,
ymax=1,
xmajorgrids,
ymajorgrids,
legend style={at={(1,0.)}, anchor=south east, legend cell align=left, align=left, draw=white!15!black,font=\small},
xlabel=$P_{\text{fa}}$,
ylabel=$P_\text{d}$,
mark size=1.5pt
]
\addplot [color=mycolor1,line width=1.2pt,mark=*,mark repeat=40]
  table[row sep=crcr]{%
1	1\\
1	1\\
1	1\\
1	1\\
1	1\\
1	1\\
1	1\\
0.9999	0.9999\\
0.9999	0.9999\\
0.9998	0.9999\\
0.9995	0.9999\\
0.9994	0.9998\\
0.9992	0.9997\\
0.9984	0.9994\\
0.997	0.9992\\
0.9945	0.9984\\
0.9922	0.9978\\
0.9883	0.9958\\
0.9839	0.9942\\
0.979	0.9924\\
0.9719	0.9897\\
0.9639	0.9871\\
0.9564	0.9842\\
0.9471	0.9805\\
0.9366	0.9758\\
0.9239	0.9709\\
0.9124	0.966\\
0.9013	0.9611\\
0.8891	0.9572\\
0.8775	0.9516\\
0.8645	0.9451\\
0.8514	0.9385\\
0.8377	0.9296\\
0.8223	0.9214\\
0.809	0.9153\\
0.7938	0.9077\\
0.7782	0.9008\\
0.7652	0.8931\\
0.7514	0.8843\\
0.7361	0.8752\\
0.7201	0.8673\\
0.7053	0.8591\\
0.6885	0.8512\\
0.6742	0.8416\\
0.6607	0.8336\\
0.6458	0.8246\\
0.6307	0.8153\\
0.6168	0.8075\\
0.6009	0.798\\
0.5864	0.7878\\
0.5731	0.7808\\
0.558	0.7718\\
0.5434	0.764\\
0.5321	0.756\\
0.5189	0.7485\\
0.5054	0.7397\\
0.4936	0.7306\\
0.4826	0.7217\\
0.472	0.7133\\
0.4606	0.7061\\
0.4495	0.6973\\
0.4384	0.688\\
0.4276	0.6816\\
0.4167	0.6724\\
0.4075	0.6644\\
0.3981	0.6558\\
0.3889	0.6484\\
0.3811	0.6396\\
0.3726	0.6316\\
0.3648	0.6234\\
0.3558	0.6155\\
0.3473	0.6065\\
0.3378	0.5977\\
0.3294	0.5876\\
0.3214	0.5805\\
0.3132	0.5722\\
0.3058	0.5638\\
0.2981	0.5564\\
0.2903	0.5498\\
0.2823	0.5429\\
0.2749	0.5352\\
0.2672	0.527\\
0.2611	0.521\\
0.2554	0.5127\\
0.2502	0.5057\\
0.2424	0.4987\\
0.2369	0.4916\\
0.2314	0.4845\\
0.225	0.478\\
0.2186	0.472\\
0.2143	0.4674\\
0.21	0.4621\\
0.2053	0.4566\\
0.201	0.449\\
0.197	0.4421\\
0.1921	0.4364\\
0.1874	0.4299\\
0.1813	0.4231\\
0.1775	0.4165\\
0.1723	0.4107\\
0.1686	0.4046\\
0.1655	0.3992\\
0.1607	0.3924\\
0.1566	0.3872\\
0.1533	0.3826\\
0.1502	0.3764\\
0.1454	0.3719\\
0.1423	0.3662\\
0.1395	0.3602\\
0.1358	0.3549\\
0.1328	0.35\\
0.1296	0.3458\\
0.1272	0.34\\
0.124	0.3355\\
0.121	0.3301\\
0.1173	0.3256\\
0.1148	0.3215\\
0.112	0.3174\\
0.11	0.3132\\
0.1079	0.3081\\
0.1055	0.3053\\
0.1031	0.3017\\
0.1	0.2984\\
0.0967	0.2954\\
0.0948	0.2913\\
0.0924	0.2874\\
0.09	0.2831\\
0.0881	0.2792\\
0.0866	0.2766\\
0.0845	0.2724\\
0.0821	0.2684\\
0.0806	0.2651\\
0.0785	0.2622\\
0.0769	0.2592\\
0.0751	0.2565\\
0.0735	0.2535\\
0.0718	0.2507\\
0.0701	0.2472\\
0.0687	0.2434\\
0.0681	0.2395\\
0.0658	0.2367\\
0.0638	0.2326\\
0.0626	0.2291\\
0.0614	0.2253\\
0.0603	0.2225\\
0.0589	0.2197\\
0.0573	0.2163\\
0.0554	0.2132\\
0.0543	0.2105\\
0.0533	0.2074\\
0.0523	0.2054\\
0.0515	0.2029\\
0.0505	0.1999\\
0.0493	0.1967\\
0.0482	0.194\\
0.0471	0.1908\\
0.0461	0.1885\\
0.0454	0.1859\\
0.044	0.1829\\
0.0432	0.1801\\
0.0427	0.1779\\
0.0418	0.1756\\
0.0411	0.1728\\
0.0406	0.1705\\
0.0397	0.1681\\
0.0388	0.1654\\
0.0378	0.1629\\
0.0374	0.1608\\
0.0367	0.1581\\
0.0361	0.1555\\
0.0349	0.1536\\
0.034	0.1514\\
0.0334	0.1497\\
0.033	0.1482\\
0.0324	0.1456\\
0.0315	0.144\\
0.0308	0.1419\\
0.0298	0.1401\\
0.0292	0.1385\\
0.0286	0.1367\\
0.0278	0.135\\
0.0271	0.1332\\
0.0268	0.1311\\
0.0257	0.1297\\
0.0253	0.1288\\
0.025	0.1269\\
0.0246	0.1256\\
0.0242	0.124\\
0.0238	0.1225\\
0.0232	0.1206\\
0.023	0.119\\
0.0225	0.1182\\
0.0217	0.1171\\
0.0215	0.1154\\
0.0211	0.1141\\
0.0208	0.1135\\
0.0205	0.1123\\
0.02	0.1104\\
0.0199	0.1098\\
0.0193	0.1081\\
0.0191	0.1068\\
0.0186	0.1059\\
0.0181	0.1049\\
0.0179	0.1039\\
0.0177	0.1022\\
0.0175	0.1007\\
0.0169	0.0998\\
0.0164	0.0984\\
0.016	0.0973\\
0.0158	0.0966\\
0.0152	0.0955\\
0.0148	0.0944\\
0.0144	0.0935\\
0.0141	0.0917\\
0.0141	0.0909\\
0.0138	0.0894\\
0.0135	0.0885\\
0.0133	0.0875\\
0.013	0.0868\\
0.0129	0.086\\
0.0128	0.0852\\
0.0127	0.0843\\
0.0124	0.0834\\
0.0122	0.083\\
0.012	0.0819\\
0.0118	0.0814\\
0.0116	0.0802\\
0.0114	0.0784\\
0.0111	0.0773\\
0.0109	0.0762\\
0.0109	0.0753\\
0.0109	0.074\\
0.0105	0.0728\\
0.0105	0.072\\
0.0104	0.0715\\
0.0104	0.0709\\
0.0103	0.07\\
0.0103	0.0693\\
0.01	0.0685\\
0.0099	0.068\\
0.0097	0.0676\\
0.0096	0.0665\\
0.0096	0.0657\\
0.0092	0.0649\\
0.0091	0.0641\\
0.0088	0.0632\\
0.0088	0.0625\\
0.0087	0.0615\\
0.0085	0.0611\\
0.0084	0.0605\\
0.0084	0.0596\\
0.0083	0.0587\\
0.0082	0.058\\
0.0078	0.0573\\
0.0077	0.0563\\
0.0077	0.056\\
0.0077	0.0554\\
0.0076	0.0549\\
0.0075	0.0541\\
0.0073	0.0531\\
0.0072	0.0524\\
0.0071	0.0522\\
0.0069	0.0519\\
0.0069	0.0513\\
0.0069	0.0511\\
0.0068	0.0507\\
0.0067	0.0501\\
0.0065	0.0497\\
0.0063	0.0494\\
0.0063	0.049\\
0.0062	0.0488\\
0.0061	0.0483\\
0.006	0.0477\\
0.006	0.0473\\
0.0059	0.047\\
0.0059	0.0463\\
0.0057	0.0459\\
0.0056	0.0455\\
0.0055	0.0444\\
0.0055	0.0438\\
0.0053	0.043\\
0.0053	0.0426\\
0.0053	0.0423\\
0.0052	0.0419\\
0.0052	0.0414\\
0.0051	0.0413\\
0.0051	0.0406\\
0.005	0.0401\\
0.005	0.0395\\
0.005	0.0393\\
0.0049	0.0387\\
0.0049	0.0384\\
0.0049	0.0383\\
0.0049	0.0381\\
0.0048	0.0372\\
0.0048	0.0367\\
0.0046	0.0363\\
0.0046	0.0358\\
};
\addlegendentry{theorectical}

\addplot [color=mycolor4, line width=1.2pt,mark=triangle*,mark repeat=55]
table[row sep=crcr]{%
	1	1\\
	1	1\\
	0.9999	0.9999\\
	0.9997	0.9999\\
	0.9991	0.9997\\
	0.9979	0.9995\\
	0.9953	0.9991\\
	0.9921	0.9972\\
	0.9865	0.9955\\
	0.9798	0.993\\
	0.9713	0.9908\\
	0.9609	0.9868\\
	0.9489	0.9823\\
	0.9371	0.9777\\
	0.924	0.9726\\
	0.9102	0.9663\\
	0.8958	0.9602\\
	0.8825	0.9544\\
	0.8697	0.947\\
	0.8559	0.9402\\
	0.8426	0.9338\\
	0.8254	0.9258\\
	0.8114	0.9199\\
	0.7965	0.9111\\
	0.7808	0.9037\\
	0.7649	0.8947\\
	0.7478	0.8863\\
	0.732	0.8785\\
	0.7173	0.8689\\
	0.701	0.8588\\
	0.684	0.8492\\
	0.6681	0.8401\\
	0.6523	0.833\\
	0.6372	0.8247\\
	0.6261	0.8162\\
	0.6127	0.8063\\
	0.5984	0.7975\\
	0.5842	0.7888\\
	0.5685	0.7811\\
	0.5539	0.7721\\
	0.5398	0.7633\\
	0.5247	0.7543\\
	0.5103	0.7448\\
	0.4963	0.7362\\
	0.4843	0.7289\\
	0.474	0.7193\\
	0.4631	0.7095\\
	0.454	0.702\\
	0.4431	0.6955\\
	0.4311	0.6879\\
	0.4197	0.6777\\
	0.4097	0.6688\\
	0.4003	0.6585\\
	0.3908	0.6508\\
	0.3814	0.6431\\
	0.3726	0.6344\\
	0.363	0.6275\\
	0.3541	0.6175\\
	0.3452	0.6082\\
	0.3373	0.6003\\
	0.3277	0.5922\\
	0.3196	0.5832\\
	0.3127	0.5751\\
	0.307	0.567\\
	0.2987	0.5593\\
	0.2915	0.552\\
	0.2837	0.5451\\
	0.2763	0.5378\\
	0.2684	0.5302\\
	0.2615	0.5209\\
	0.2545	0.5129\\
	0.2479	0.5058\\
	0.2404	0.4996\\
	0.235	0.4923\\
	0.2289	0.4861\\
	0.2233	0.4798\\
	0.2174	0.4735\\
	0.2124	0.4665\\
	0.2071	0.4615\\
	0.2025	0.4563\\
	0.1983	0.4509\\
	0.1944	0.4465\\
	0.1895	0.4403\\
	0.1854	0.4333\\
	0.1801	0.4268\\
	0.1742	0.4201\\
	0.1703	0.4128\\
	0.1666	0.4078\\
	0.1629	0.3994\\
	0.1598	0.3942\\
	0.1566	0.3867\\
	0.1525	0.3817\\
	0.1489	0.3772\\
	0.1452	0.3709\\
	0.1408	0.3655\\
	0.1385	0.3593\\
	0.1345	0.3543\\
	0.1307	0.35\\
	0.1275	0.3443\\
	0.1247	0.3385\\
	0.122	0.3346\\
	0.1182	0.3301\\
	0.1151	0.3252\\
	0.1134	0.3206\\
	0.1107	0.3177\\
	0.1081	0.3138\\
	0.1051	0.3098\\
	0.1029	0.3068\\
	0.1005	0.3018\\
	0.0982	0.2976\\
	0.0959	0.295\\
	0.0929	0.2905\\
	0.0907	0.2868\\
	0.0892	0.2823\\
	0.0875	0.2774\\
	0.0859	0.2743\\
	0.0838	0.2715\\
	0.0812	0.2672\\
	0.0797	0.2636\\
	0.0774	0.2604\\
	0.076	0.2574\\
	0.0747	0.2545\\
	0.0735	0.2519\\
	0.0714	0.2488\\
	0.0685	0.2455\\
	0.0666	0.2409\\
	0.0653	0.2387\\
	0.0638	0.235\\
	0.0623	0.2304\\
	0.0612	0.2277\\
	0.0599	0.2252\\
	0.0584	0.2221\\
	0.0568	0.2192\\
	0.0555	0.216\\
	0.0539	0.2127\\
	0.0527	0.2089\\
	0.0515	0.2061\\
	0.0501	0.2036\\
	0.0495	0.2001\\
	0.0485	0.198\\
	0.0477	0.195\\
	0.0465	0.1925\\
	0.0458	0.1903\\
	0.0452	0.1882\\
	0.0446	0.1859\\
	0.0436	0.1836\\
	0.0429	0.1803\\
	0.042	0.1781\\
	0.0411	0.176\\
	0.0402	0.1726\\
	0.04	0.1696\\
	0.0395	0.1675\\
	0.0384	0.1649\\
	0.0373	0.1617\\
	0.0361	0.1595\\
	0.0353	0.1569\\
	0.0345	0.1545\\
	0.0336	0.1521\\
	0.0328	0.1499\\
	0.0321	0.148\\
	0.0312	0.1468\\
	0.0306	0.145\\
	0.0301	0.1432\\
	0.0294	0.1418\\
	0.0285	0.1406\\
	0.028	0.1389\\
	0.0278	0.1366\\
	0.0274	0.1343\\
	0.0266	0.1327\\
	0.0265	0.1308\\
	0.0257	0.1296\\
	0.025	0.1278\\
	0.0242	0.1258\\
	0.024	0.1242\\
	0.0236	0.1233\\
	0.0234	0.1213\\
	0.0231	0.1201\\
	0.0226	0.1188\\
	0.0221	0.1172\\
	0.022	0.1159\\
	0.0217	0.1143\\
	0.0212	0.1131\\
	0.0209	0.1127\\
	0.0207	0.1114\\
	0.0205	0.1094\\
	0.0201	0.1088\\
	0.0196	0.1078\\
	0.0188	0.1066\\
	0.0185	0.1054\\
	0.0181	0.1043\\
	0.0181	0.1028\\
	0.0173	0.1017\\
	0.0166	0.1006\\
	0.0163	0.0988\\
	0.0158	0.0978\\
	0.0152	0.0964\\
	0.015	0.0953\\
	0.0148	0.0951\\
	0.0145	0.0934\\
	0.0144	0.0924\\
	0.014	0.091\\
	0.0138	0.09\\
	0.0137	0.0893\\
	0.0134	0.088\\
	0.0133	0.0875\\
	0.0129	0.0862\\
	0.0126	0.0853\\
	0.0123	0.0842\\
	0.0121	0.0834\\
	0.0119	0.083\\
	0.0118	0.0819\\
	0.0116	0.0804\\
	0.0113	0.0797\\
	0.0111	0.0787\\
	0.0111	0.0778\\
	0.011	0.0768\\
	0.0107	0.0759\\
	0.0107	0.0752\\
	0.0107	0.0739\\
	0.0106	0.0731\\
	0.0106	0.072\\
	0.0105	0.0706\\
	0.0104	0.07\\
	0.0102	0.0692\\
	0.0099	0.0689\\
	0.0098	0.0682\\
	0.0098	0.0674\\
	0.0097	0.0667\\
	0.0096	0.0658\\
	0.0095	0.0649\\
	0.0092	0.0641\\
	0.0087	0.0629\\
	0.0087	0.0623\\
	0.0085	0.0613\\
	0.0082	0.0605\\
	0.0082	0.0602\\
	0.0081	0.0595\\
	0.0081	0.0588\\
	0.0081	0.0581\\
	0.0079	0.0573\\
	0.0078	0.0565\\
	0.0077	0.0557\\
	0.0077	0.0552\\
	0.0074	0.0549\\
	0.0073	0.0542\\
	0.0073	0.0538\\
	0.0073	0.0529\\
	0.0072	0.0526\\
	0.007	0.0521\\
	0.007	0.0516\\
	0.0068	0.0509\\
	0.0067	0.0505\\
	0.0067	0.0503\\
	0.0065	0.0499\\
	0.0064	0.0496\\
	0.0064	0.0495\\
	0.0061	0.0492\\
	0.006	0.0482\\
	0.006	0.0474\\
	0.0059	0.0469\\
	0.0059	0.0464\\
	0.0058	0.0462\\
	0.0057	0.0459\\
	0.0056	0.0457\\
	0.0056	0.0451\\
	0.0054	0.0444\\
	0.0053	0.044\\
	0.0052	0.0435\\
	0.0052	0.043\\
	0.0052	0.0424\\
	0.0052	0.0417\\
	0.0052	0.0412\\
	0.0051	0.0408\\
	0.0051	0.0404\\
	0.0051	0.0399\\
	0.0051	0.0396\\
	0.005	0.0391\\
	0.0049	0.0389\\
	0.0048	0.0382\\
	0.0048	0.0381\\
	0.0047	0.0377\\
	0.0047	0.0372\\
	0.0046	0.0363\\
	0.0045	0.0359\\
	0.0044	0.0356\\
	0.0044	0.0351\\
	0.0044	0.0345\\
	0.0043	0.0339\\
	0.0042	0.0337\\
	0.0042	0.0336\\
	0.0042	0.0335\\
	0.004	0.0327\\
	0.0039	0.0323\\
	0.0037	0.0317\\
	0.0036	0.0313\\
	0.0036	0.0307\\
	0.0036	0.0304\\
	0.0036	0.03\\
};
\addlegendentry{d-pmSVD2}

\addplot [color=mycolor2, line width=1.2pt,mark=square*,mark repeat=45]
  table[row sep=crcr]{%
1	1\\
1	1\\
1	1\\
1	1\\
1	1\\
1	1\\
1	1\\
0.9999	0.9999\\
0.9999	0.9999\\
0.9998	0.9999\\
0.9995	0.9999\\
0.9994	0.9998\\
0.9992	0.9997\\
0.9984	0.9994\\
0.997	0.9992\\
0.9945	0.9984\\
0.9922	0.9978\\
0.9883	0.9958\\
0.9839	0.9942\\
0.979	0.9924\\
0.9719	0.9897\\
0.9639	0.9871\\
0.9564	0.9842\\
0.9471	0.9805\\
0.9366	0.9758\\
0.9239	0.9709\\
0.9124	0.966\\
0.9013	0.9611\\
0.8891	0.9572\\
0.8775	0.9516\\
0.8645	0.9451\\
0.8514	0.9385\\
0.8377	0.9296\\
0.8223	0.9214\\
0.809	0.9153\\
0.7938	0.9077\\
0.7782	0.9008\\
0.7652	0.8931\\
0.7514	0.8843\\
0.7361	0.8752\\
0.7201	0.8673\\
0.7053	0.8591\\
0.6885	0.8512\\
0.6742	0.8416\\
0.6607	0.8336\\
0.6458	0.8246\\
0.6307	0.8153\\
0.6168	0.8075\\
0.6009	0.798\\
0.5864	0.7878\\
0.5731	0.7808\\
0.558	0.7718\\
0.5434	0.764\\
0.5321	0.756\\
0.5189	0.7485\\
0.5054	0.7397\\
0.4936	0.7306\\
0.4826	0.7217\\
0.472	0.7133\\
0.4606	0.7061\\
0.4495	0.6973\\
0.4384	0.688\\
0.4276	0.6816\\
0.4167	0.6724\\
0.4075	0.6644\\
0.3981	0.6558\\
0.3889	0.6484\\
0.3811	0.6396\\
0.3726	0.6316\\
0.3648	0.6234\\
0.3558	0.6155\\
0.3473	0.6065\\
0.3378	0.5977\\
0.3294	0.5876\\
0.3214	0.5805\\
0.3132	0.5722\\
0.3058	0.5638\\
0.2981	0.5564\\
0.2903	0.5498\\
0.2823	0.5429\\
0.2749	0.5352\\
0.2672	0.527\\
0.2611	0.521\\
0.2554	0.5127\\
0.2502	0.5057\\
0.2424	0.4987\\
0.2369	0.4916\\
0.2314	0.4845\\
0.225	0.478\\
0.2186	0.472\\
0.2143	0.4674\\
0.21	0.4621\\
0.2053	0.4566\\
0.201	0.449\\
0.197	0.4421\\
0.1921	0.4364\\
0.1874	0.4299\\
0.1813	0.4231\\
0.1775	0.4165\\
0.1723	0.4107\\
0.1686	0.4046\\
0.1655	0.3992\\
0.1607	0.3924\\
0.1566	0.3872\\
0.1533	0.3826\\
0.1502	0.3764\\
0.1454	0.3719\\
0.1423	0.3662\\
0.1395	0.3602\\
0.1358	0.3549\\
0.1328	0.35\\
0.1296	0.3458\\
0.1272	0.34\\
0.124	0.3355\\
0.121	0.3301\\
0.1173	0.3256\\
0.1148	0.3215\\
0.112	0.3174\\
0.11	0.3132\\
0.1079	0.3081\\
0.1055	0.3053\\
0.1031	0.3017\\
0.1	0.2984\\
0.0967	0.2954\\
0.0948	0.2913\\
0.0924	0.2874\\
0.09	0.2831\\
0.0881	0.2792\\
0.0866	0.2766\\
0.0845	0.2724\\
0.0821	0.2684\\
0.0806	0.2651\\
0.0785	0.2622\\
0.0769	0.2592\\
0.0751	0.2565\\
0.0735	0.2535\\
0.0718	0.2507\\
0.0701	0.2472\\
0.0687	0.2434\\
0.0681	0.2395\\
0.0658	0.2367\\
0.0638	0.2326\\
0.0626	0.2291\\
0.0614	0.2253\\
0.0603	0.2225\\
0.0589	0.2197\\
0.0573	0.2163\\
0.0554	0.2132\\
0.0543	0.2105\\
0.0533	0.2074\\
0.0523	0.2054\\
0.0515	0.2029\\
0.0505	0.1999\\
0.0493	0.1967\\
0.0482	0.194\\
0.0471	0.1908\\
0.0461	0.1885\\
0.0454	0.1859\\
0.044	0.1829\\
0.0432	0.1801\\
0.0427	0.1779\\
0.0418	0.1756\\
0.0411	0.1728\\
0.0406	0.1705\\
0.0397	0.1681\\
0.0388	0.1654\\
0.0378	0.1629\\
0.0374	0.1608\\
0.0367	0.1581\\
0.0361	0.1555\\
0.0349	0.1536\\
0.034	0.1514\\
0.0334	0.1497\\
0.033	0.1482\\
0.0324	0.1456\\
0.0315	0.144\\
0.0308	0.1419\\
0.0298	0.1401\\
0.0292	0.1385\\
0.0286	0.1367\\
0.0278	0.135\\
0.0271	0.1332\\
0.0268	0.1311\\
0.0257	0.1297\\
0.0253	0.1288\\
0.025	0.1269\\
0.0246	0.1256\\
0.0242	0.124\\
0.0238	0.1225\\
0.0232	0.1206\\
0.023	0.119\\
0.0225	0.1182\\
0.0217	0.1171\\
0.0215	0.1154\\
0.0211	0.1141\\
0.0208	0.1135\\
0.0205	0.1123\\
0.02	0.1104\\
0.0199	0.1098\\
0.0193	0.1081\\
0.0191	0.1068\\
0.0186	0.1059\\
0.0181	0.1049\\
0.0179	0.1039\\
0.0177	0.1022\\
0.0175	0.1007\\
0.0169	0.0998\\
0.0164	0.0984\\
0.016	0.0973\\
0.0158	0.0966\\
0.0152	0.0955\\
0.0148	0.0944\\
0.0144	0.0935\\
0.0141	0.0917\\
0.0141	0.0909\\
0.0138	0.0894\\
0.0135	0.0885\\
0.0133	0.0875\\
0.013	0.0868\\
0.0129	0.086\\
0.0128	0.0852\\
0.0127	0.0843\\
0.0124	0.0834\\
0.0122	0.083\\
0.012	0.0819\\
0.0118	0.0814\\
0.0116	0.0802\\
0.0114	0.0784\\
0.0111	0.0773\\
0.0109	0.0762\\
0.0109	0.0753\\
0.0109	0.074\\
0.0105	0.0728\\
0.0105	0.072\\
0.0104	0.0715\\
0.0104	0.0709\\
0.0103	0.07\\
0.0103	0.0693\\
0.01	0.0685\\
0.0099	0.068\\
0.0097	0.0676\\
0.0096	0.0665\\
0.0096	0.0657\\
0.0092	0.0649\\
0.0091	0.0641\\
0.0088	0.0632\\
0.0088	0.0625\\
0.0087	0.0615\\
0.0085	0.0611\\
0.0084	0.0605\\
0.0084	0.0596\\
0.0083	0.0587\\
0.0082	0.058\\
0.0078	0.0573\\
0.0077	0.0563\\
0.0077	0.056\\
0.0077	0.0554\\
0.0076	0.0549\\
0.0075	0.0541\\
0.0073	0.0531\\
0.0072	0.0524\\
0.0071	0.0522\\
0.0069	0.0519\\
0.0069	0.0513\\
0.0069	0.0511\\
0.0068	0.0507\\
0.0067	0.0501\\
0.0065	0.0497\\
0.0063	0.0494\\
0.0063	0.049\\
0.0062	0.0488\\
0.0061	0.0483\\
0.006	0.0477\\
0.006	0.0473\\
0.0059	0.047\\
0.0059	0.0463\\
0.0057	0.0459\\
0.0056	0.0455\\
0.0055	0.0444\\
0.0055	0.0438\\
0.0053	0.043\\
0.0053	0.0426\\
0.0053	0.0423\\
0.0052	0.0419\\
0.0052	0.0414\\
0.0051	0.0413\\
0.0051	0.0406\\
0.005	0.0401\\
0.005	0.0395\\
0.005	0.0393\\
0.0049	0.0387\\
0.0049	0.0384\\
0.0049	0.0383\\
0.0049	0.0381\\
0.0048	0.0372\\
0.0048	0.0367\\
0.0046	0.0363\\
0.0046	0.0358\\
};
\addlegendentry{d-raSVD2}

\addplot [color=mycolor2, dashdotted, line width=1.2pt]
table[row sep=crcr]{%
	1	1\\
	1	1\\
	1	1\\
	1	1\\
	0.9999	0.9999\\
	0.9997	0.9999\\
	0.9994	0.9998\\
	0.9985	0.9997\\
	0.9958	0.9991\\
	0.9928	0.998\\
	0.9873	0.9953\\
	0.9811	0.9929\\
	0.972	0.99\\
	0.9631	0.9871\\
	0.9511	0.9827\\
	0.9374	0.9779\\
	0.9248	0.9721\\
	0.9113	0.9674\\
	0.8971	0.9611\\
	0.8833	0.9549\\
	0.8704	0.9487\\
	0.8549	0.9415\\
	0.8411	0.9345\\
	0.8271	0.9263\\
	0.8098	0.9188\\
	0.7943	0.9119\\
	0.7806	0.9039\\
	0.7646	0.894\\
	0.7495	0.8852\\
	0.7332	0.8771\\
	0.7162	0.8683\\
	0.7006	0.8587\\
	0.6845	0.8486\\
	0.6706	0.8401\\
	0.6543	0.8322\\
	0.6398	0.8246\\
	0.6256	0.8159\\
	0.6116	0.8072\\
	0.5978	0.7974\\
	0.5849	0.7899\\
	0.5701	0.7816\\
	0.556	0.7731\\
	0.5415	0.7656\\
	0.5273	0.7546\\
	0.5135	0.7456\\
	0.5001	0.738\\
	0.4873	0.7292\\
	0.4747	0.7214\\
	0.4652	0.7119\\
	0.4522	0.7026\\
	0.4433	0.6948\\
	0.4328	0.6882\\
	0.4219	0.6801\\
	0.4106	0.6709\\
	0.4006	0.6628\\
	0.3923	0.6537\\
	0.3833	0.6447\\
	0.3738	0.6358\\
	0.3653	0.6287\\
	0.3574	0.6198\\
	0.3477	0.6118\\
	0.3385	0.6022\\
	0.3298	0.5944\\
	0.3219	0.5864\\
	0.314	0.5774\\
	0.3065	0.57\\
	0.2996	0.5606\\
	0.292	0.5531\\
	0.2834	0.5455\\
	0.2768	0.5378\\
	0.2698	0.5308\\
	0.2627	0.5227\\
	0.2546	0.5149\\
	0.2485	0.5073\\
	0.243	0.5006\\
	0.2364	0.4947\\
	0.2308	0.4881\\
	0.225	0.4809\\
	0.2201	0.4748\\
	0.214	0.4683\\
	0.2091	0.4621\\
	0.2044	0.4567\\
	0.2	0.4519\\
	0.1953	0.4457\\
	0.1908	0.4412\\
	0.1858	0.4348\\
	0.1814	0.4278\\
	0.1763	0.4212\\
	0.1715	0.4146\\
	0.1681	0.409\\
	0.1641	0.4027\\
	0.1604	0.3956\\
	0.1572	0.3896\\
	0.1534	0.3834\\
	0.1497	0.3791\\
	0.146	0.3725\\
	0.1419	0.3672\\
	0.1382	0.3629\\
	0.1349	0.357\\
	0.1316	0.3508\\
	0.1286	0.3458\\
	0.1259	0.3399\\
	0.1222	0.3347\\
	0.1192	0.331\\
	0.1163	0.3269\\
	0.1136	0.3225\\
	0.1107	0.3176\\
	0.109	0.3141\\
	0.1064	0.3107\\
	0.1032	0.3077\\
	0.1001	0.3034\\
	0.0983	0.2993\\
	0.0962	0.2959\\
	0.0934	0.2922\\
	0.0912	0.2882\\
	0.0891	0.2839\\
	0.0875	0.28\\
	0.0858	0.2758\\
	0.0841	0.273\\
	0.082	0.2705\\
	0.0796	0.2661\\
	0.0784	0.2628\\
	0.0767	0.2583\\
	0.0746	0.2549\\
	0.0729	0.2519\\
	0.0708	0.2496\\
	0.0692	0.2466\\
	0.0676	0.2425\\
	0.0659	0.2387\\
	0.0648	0.2364\\
	0.0636	0.2326\\
	0.0617	0.2294\\
	0.0604	0.225\\
	0.0591	0.2218\\
	0.0576	0.2184\\
	0.0562	0.2149\\
	0.0544	0.2123\\
	0.0533	0.2105\\
	0.0518	0.2064\\
	0.0509	0.2037\\
	0.0499	0.201\\
	0.0489	0.1979\\
	0.0475	0.1954\\
	0.0463	0.1923\\
	0.0456	0.1901\\
	0.0453	0.1884\\
	0.0448	0.1865\\
	0.0438	0.1838\\
	0.0428	0.1811\\
	0.0421	0.1782\\
	0.0412	0.176\\
	0.0406	0.1727\\
	0.0399	0.1705\\
	0.0389	0.1678\\
	0.0388	0.1662\\
	0.0377	0.1635\\
	0.0368	0.161\\
	0.0358	0.159\\
	0.035	0.1562\\
	0.0338	0.153\\
	0.033	0.1508\\
	0.0323	0.1492\\
	0.0312	0.1466\\
	0.0306	0.145\\
	0.0299	0.1436\\
	0.0295	0.1422\\
	0.0287	0.1407\\
	0.0282	0.1393\\
	0.0278	0.1372\\
	0.0271	0.1359\\
	0.0268	0.1337\\
	0.0261	0.1321\\
	0.0258	0.1304\\
	0.0252	0.129\\
	0.0245	0.1271\\
	0.0243	0.1257\\
	0.0239	0.1236\\
	0.0234	0.1225\\
	0.0229	0.1213\\
	0.0223	0.1196\\
	0.0222	0.1181\\
	0.022	0.1169\\
	0.0218	0.1155\\
	0.0215	0.1142\\
	0.0208	0.113\\
	0.0204	0.1112\\
	0.0203	0.1107\\
	0.0203	0.1093\\
	0.0201	0.1076\\
	0.0193	0.1069\\
	0.0186	0.1052\\
	0.0184	0.1043\\
	0.0181	0.1034\\
	0.0175	0.1014\\
	0.017	0.1003\\
	0.0168	0.0988\\
	0.0164	0.0981\\
	0.0162	0.0972\\
	0.0152	0.0959\\
	0.0149	0.0948\\
	0.0147	0.0939\\
	0.0146	0.0927\\
	0.0144	0.0922\\
	0.0139	0.0909\\
	0.0138	0.0895\\
	0.0136	0.0887\\
	0.0134	0.0878\\
	0.013	0.087\\
	0.0125	0.0862\\
	0.0124	0.0848\\
	0.0123	0.0839\\
	0.0123	0.0826\\
	0.0119	0.0821\\
	0.0118	0.0813\\
	0.0115	0.0803\\
	0.0113	0.0792\\
	0.011	0.0777\\
	0.011	0.0765\\
	0.0108	0.0754\\
	0.0108	0.0745\\
	0.0108	0.0738\\
	0.0107	0.0731\\
	0.0106	0.072\\
	0.0104	0.0712\\
	0.0104	0.0703\\
	0.0104	0.0697\\
	0.0103	0.0686\\
	0.0101	0.0682\\
	0.01	0.0674\\
	0.0099	0.0669\\
	0.0097	0.0659\\
	0.0094	0.0649\\
	0.0091	0.0639\\
	0.0089	0.0631\\
	0.0089	0.0621\\
	0.0085	0.0616\\
	0.0082	0.0613\\
	0.0082	0.0604\\
	0.0082	0.06\\
	0.0079	0.0594\\
	0.0079	0.0583\\
	0.0077	0.0578\\
	0.0077	0.0571\\
	0.0076	0.0562\\
	0.0075	0.0556\\
	0.0074	0.0548\\
	0.0073	0.0545\\
	0.0072	0.054\\
	0.0071	0.0534\\
	0.007	0.0527\\
	0.007	0.0522\\
	0.0069	0.052\\
	0.0068	0.0513\\
	0.0068	0.0509\\
	0.0068	0.0505\\
	0.0067	0.0502\\
	0.0067	0.0496\\
	0.0066	0.0495\\
	0.0065	0.0492\\
	0.0064	0.0485\\
	0.0061	0.0478\\
	0.006	0.0471\\
	0.006	0.0463\\
	0.006	0.046\\
	0.0058	0.046\\
	0.0058	0.0456\\
	0.0056	0.0454\\
	0.0055	0.0449\\
	0.0054	0.0445\\
	0.0053	0.0437\\
	0.0053	0.0433\\
	0.0052	0.0427\\
	0.0052	0.0419\\
	0.0052	0.0415\\
	0.0052	0.0412\\
	0.0052	0.0406\\
	0.0052	0.0397\\
	0.0051	0.0396\\
	0.0049	0.0393\\
	0.0049	0.039\\
	0.0048	0.0386\\
	0.0047	0.0381\\
	0.0046	0.038\\
	0.0046	0.0371\\
	0.0045	0.0366\\
	0.0045	0.0359\\
	0.0043	0.0358\\
	0.0042	0.0353\\
	0.0042	0.0348\\
	0.0041	0.0346\\
	0.004	0.0341\\
	0.004	0.0336\\
	0.004	0.0333\\
	0.0039	0.0326\\
	0.0039	0.032\\
	0.0038	0.0313\\
	0.0037	0.0311\\
	0.0037	0.0309\\
	0.0036	0.0305\\
	0.0036	0.03\\
	0.0036	0.0297\\
};
\addlegendentry{d-TraSVD2($1$)}

\end{axis}

\end{tikzpicture}%

%% file: appendixRA.tex
\section{Rational Function Approximation Approach}\label{sec:appendixRA}
The RA approach is established by the following theorem\footnote{The iteration index $t$ is omitted here for simplicity of presentation.}:
\begin{theorem}\label{the:rank1}
	Suppose $\boldsymbol{\Lambda} = \diag{\lambda_1,\ldots,\lambda_N}\in\mathbb{R}^{N\times N}$ {where the diagonal entries are distinct and are sorted in descending order, i.e., }  $\lambda_1>\cdots>\lambda_N$. {Further assume} that $\rho\neq 0$ and $\bz = \left[z_1, \ldots, z_N\right] \in\mathbb{C}^{N}$ with $z_i\neq 0$ for all $i = 1, \ldots, N$. If $\bW = [\bw_1,\cdots,\bw_N]\in\mathbb{C}^{N\times N}$ is {an orthogonal matrix} such that
	$$
	\bW^\tH(\boldsymbol{\Lambda}+\rho \bz\bz^\tH)\bW = \diag{\bar{\lambda}_1,\ldots,\bar{\lambda}_N},
	$$
	with $\bar{\lambda}_1>\cdots>\bar{\lambda}_N$, then
	\begin{enumerate}
		\item The values in set $\{ \bar{\lambda}_i\}_{i=1}^N$ are the $N$ zeros of the secular function $f(\lambda) = 1 + \rho \bz^\tH(\boldsymbol{\Lambda}-\lambda\bI)^{-1}\bz$.
		\item The values $\{ \bar{\lambda}_i\}_{i=1}^N$ satisfy the interlacing property, i.e.,\\
		$\bar{\lambda}_1 \geq \lambda_1 \geq \bar{\lambda}_2 \geq  \cdots  \geq  \bar{\lambda}_N \geq \lambda_N$, if $\rho > 0$,\\
		$\lambda_1 \geq \bar{\lambda}_1 \geq \lambda_2 \geq \cdots  \geq  \lambda_N \geq  \bar{\lambda}_N$, if $\rho < 0$.
		\item The eigenvector $\bw_i$ associated with $\bar{\lambda}_i$ is a multiple of $(\boldsymbol{\Lambda} -\bar{\lambda}_i \bI)^{-1}\bz$.
	\end{enumerate}
\end{theorem}

Without loss of generality we assume that $\rho > 0$, otherwise, we can replace $\lambda_i$ by $-\lambda_{N-i+1}$ and $\rho$ by $-\rho$ \cite{bunchRankonemodification1978}.
According to Theorem \ref{the:rank1}.1, the eigenvalues $\bar{\lambda}_k$ of the matrix $\boldsymbol{\Lambda} + \rho \bz\bz^\tH$ can be computed by solving $f(\lambda) = 0$, i.e.,
\begin{equation}\label{equ:secular}
	f(\lambda) = 1 + \rho{\sum_{i = 1}^{N}}\frac{|z_{i}|^2}{\lambda_i-\lambda} = 0.
\end{equation}
Based on the interlacing property of the eigenvalues, for the $k$th eigenvalue $\bar{\lambda}_i\in(\lambda_{i},\lambda_{i-1})$ with $\lambda_0 = \lambda_1 + \rho \bz^\tH\bz$ \cite{liSolvingsecular1994}, we can rearrange \eqref{equ:secular} as
\begin{equation}\label{equ:ra}
	\begin{aligned}
		-\psi_{i-1}(\lambda)  & = 1 + \phi_i(\lambda),
	\end{aligned}
\end{equation}	
where\footnote{Different possible forms of the approximants, i.e., $\psi_{i-1}(\lambda)$ and $\phi_{i}(\lambda)$, and the closed-form expressions of the associated parameters are discussed and provided in \cite{liSolvingsecular1994,melmanNumericalcomparison1997}.}
\begin{equation}
	\label{equ:trueFunc}		
	\psi_{i-1}(\lambda)  = \rho {\sum_{k = 1}^{i-1}}\frac{|z_{k}|^2}{\lambda_k-\lambda}\ \text{ and }\  \phi_i(\lambda)  = \rho {\sum_{k = i}^{N}}\frac{|z_{k}|^2}{\lambda_k-\lambda}.
	\myvspace
\end{equation}
Both functions $\psi_{i-1}(\lambda)$ and $\phi_i(\lambda)$ can be approximated with simple rational functions \cite{liSolvingsecular1994,bunchRankonemodification1978} as
\begin{equation}
	\label{equ:approxFunc}
	\tilde{\psi}_{i-1}(\lambda) =\ p + \frac{q}{\lambda_{k-1} - \lambda}\ \text{ and }\  \tilde{\phi}_{i}(\lambda) =\ r + \frac{s}{\lambda_{k} - \lambda}.
\end{equation}
In \eqref{equ:approxFunc}, the parameters $p$, $q$, $r$ and $s$ are chosen such that, at the given iterate $\lambda^{(\tau)}$, the rational approximants $\tilde{\psi}_{i-1}(\lambda)$ and $\tilde{\phi}_i(\lambda)$ in \eqref{equ:approxFunc} coincide with the true rational functions $\psi_{i-1}(\lambda)$ and $\phi_{i}(\lambda)$ in \eqref{equ:trueFunc} up to the first order derivative, respectively. The next iterate $\lambda^{(\tau+1)}\in\left(\lambda_i, \lambda_{i-1}\right)$ is obtained from the solution of $-\tilde{\psi}_{i-1}(\lambda)  = 1 + \tilde{\phi}_i(\lambda)$. 
For the special case where $i = 1$, function $\psi_{i-1}(\lambda)$ is approximated as $\tilde{\psi}_0(\lambda) = 0$. The RA approach converges quadratically \cite{liSolvingsecular1994}, and requires less than $4$ iterations for a tolerance $\abs{\lambda^{(\tau+1)} - \lambda^{(\tau)}} = 10^{-9}$ \cite{trinh-hoangPartialrelaxation2020}.

%% file: appendix1.tex
\section{Resolving Sorting Order Ambiguity for Rational Function Approximation in Different Nodes}\label{sec:appendix1}
The RA approach requires the eigenvalues to be sorted descendingly, and the eigenvectors to be sorted accordingly. Nonetheless, due to machine precision limitations and the nature of distributed consensus algorithms, different nodes may obtain slightly different results of the eigenvalues, and thus the sorting order of them may differ. The following example is provided to illustrate the sorting order ambiguity and the proposed local sorting rule.


Denote the unsorted solutions of \eqref{equ:rank1modified} via the RA approach in the $i$th and the $j$th node as $\bar{\bLambda}^\prime$ and $\bar{\bLambda}$, respectively. Specifically, it is assumed that the respective diagonal entries
\begin{equation}\label{equ:sorting}
	\begin{aligned}
		\bar{\lambda}_1^\prime &= 2, &\bar{\lambda}_1 &= 2,\\
		\bar{\lambda}_2^\prime &= 1.0000001, &\bar{\lambda}_2 &= 1.0000001,\\
		\bar{\lambda}_3^\prime &= 0.9999999, &\bar{\lambda}_3 &= 0.9999999, \\
		\bar{\lambda}_4^\prime &= 0, & \bar{\lambda}_4 &= 0, \\
		\bar{\lambda}_5^\prime &= \underline{0.9999999}, &\bar{\lambda}_5 &= \underline{1},
	\end{aligned}
\end{equation}
where we have highlighted the last eigenvalue with the underline operator. The values in \eqref{equ:sorting} represent local realizations of the true eigenvalues of $\diag{2,1,1,0,1}$, in two different nodes, i.e., the $i$th and $j$th node. If we apply, e.g., the \texttt{sort()} function, directly, the sorting order in the $i$th node is $1,2,3,5,4$, while in all the other nodes (including the $j$th node) is $1,2,5,3,4$. Such non-unified sorting order among all nodes result in swapped columns in the associated eigenvector matrix $\bW^\prime$ in the $i$th node as mentioned in Section \ref{subsec:oded}.

To resolve this ambiguity of the sorting order and to avoid the extra communication cost to reach the consensus on the sorting order, we sort the eigenvalues and the eigenvectors in each node separately. Specifically, for the aforementioned special case, on the one hand, we treat the eigenvalues whose difference is less than the tolerance, e.g., $10^{-6}$, as identical. Based on this criteria, the second, the third, and the fifth eigenvalues are treated as identical in all nodes. Consequently, the sorting order of all nodes is unified as $1,2,3,5,4$, and the eigenvectors in each node can be sorted accordingly. On the other hand, since the RA approach in the next iteration requires the eigenvalues to be sorted descendingly, the eigenvalues in the $i$th node are sorted with the order $1,2,3,5,4$, e.g, by directly applying the \texttt{sort()} function. Note that this approach does not introduce mismatching of the eigenvalues and the associated eigenvectors, but manually rearranges the order of the eigenvectors associated with the same eigenvalues.